\documentclass[printer]{aa}

\usepackage{natbib}
\usepackage{amsmath}
\usepackage{graphicx}
\usepackage{txfonts}

\newcommand{\rg}{GM_{\rm BH}/c^2}
\newcommand{\tu}{GM_{\rm BH}/c^3}

\newcommand{\tratd}{(\frac{T_{\rm p}}{T_{\rm e}})_{\rm disk}}
\newcommand{\tj}{\Theta_{\rm e,jet}}
\newcommand{\msun}{M_{\rm \odot}}

\newcommand{\mdot}{\dot{M}}
\newcommand{\mdotu}{\rm M_\odot yr^{-1}}

\begin{document}

\title{Observational appearance of inefficient accretion flows and jets in 3D
  GRMHD simulations: Application to Sgr~A*.}
\titlerunning{Radiation from RIAF models}
\authorrunning{Mo{\'s}cibrodzka, Falcke, Shiokawa, Gammie}
\author{Monika Mo{\'s}cibrodzka~\inst{1}, Heino Falcke~\inst{1,2}, Hotaka
  Shiokawa\inst{3}, Charles F. Gammie\inst{4,5}}
\institute{$^1$Department of Astrophysics/IMAPP,Radboud University
  Nijmegen,P.O. Box 9010, 6500 GL Nijmegen, The Netherlands\\
$^2$ASTRON, Oude Hoogeveensedijk 4, 7991 PD, Dwingeloo, The Netherlands 
\\$^3$Department of Physics \&
  Astronomy, The Johns Hopkins University 3400 N. Charles Street
Baltimore, MD 21218\\$^4$Department of Physics,
  University of Illinois, 1110 West Green Street, Urbana, IL, 61801\\
$^5$Astronomy Department, University of Illinois, 1002 West Green Street, Urbana, IL, 61801\\
\email{m.moscibrodzka@astro.ru.nl}}
\date{Received June, 2014; accepted x, 2014}

\abstract
{ Radiatively inefficient accretion flows (RIAFs) are believed to
  power supermassive black holes in the underluminous cores of
  galaxies. Such black holes are typically accompanied by flat-spectrum
  radio cores indicating the presence of moderately relativistic
  jets. One of the best constrained RIAFs is associated with the
  supermassive black hole in the Galactic center, Sgr A*.}
    { Since the plasma in RIAFs is only weakly collisional, the
      dynamics and the radiative properties of these systems are very
      uncertain.  Here we want to study the impact of varying
      electron temperature on the appearance of accretion flows and
      jets.}
    { Using three-dimensional general relativistic
      magnetohydrodynamics accretion flow simulations, we use ray
      tracing methods to predict spectra and radio images of RIAFs
      allowing for different electron heating mechanisms in the in- 
      and outflowing parts of the simulations.}
    { We find that small changes in the electron temperature
      can result in dramatic differences in the relative
      dominance of jets and accretion flows. Application to Sgr A*
      shows that radio spectrum and size of this source can be well
      reproduced with a model where electrons are more efficiently
      heated in the jet. The X-ray emission is sensitive 
      to the electron heating mechanism in the
      jets and disk and therefore X-ray observations put strong constraints
      on electron temperatures and geometry of the accretion flow and jet.
      For Sgr~A*, the jet model also predicts a significant
      frequency-dependent core shift which could place independent constraints
      on the model once measured accurately. 
     }
    { We conclude that more sophisticated models for electron
      distribution functions are crucial for constraining GRMHD
      simulations with actual observations. For Sgr A*, the radio
      appearance may well be dominated by the outflowing
      plasma. Nonetheless, at the highest radio frequencies, the
      shadow of the event horizon should still be detectable with
      future Very Long Baseline Interferometric observations. }


   \maketitle
\section{Introduction}

Supermassive black holes exist in the centers of many galaxies. Observations
suggest that plasma flows into the deep gravitational potential of black
holes, thereby releasing enormous amounts of energy in the form of radiation
and powerful plasma outflows, or jets, that can reach way beyond the host
galaxy. Very Long Baseline Interferometry, such as the Event Horizon Telescope
 \citep{doelemanwp:2009} promises to take the first ever images of the event
horizon and the plasma flow in the immediate vicinity of black holes \citep{falcke:2000a}.
Hence, a theoretical understanding
of black hole astrophysics is now crucial and timely.

An approximately four million solar mass object detected in the central
sub-arcsecond of the Milky Way is one of the best supermassive 
black hole candidates
(\citealt{genzel:2010}, \citealt{falcke:2013}).  Interestingly, the
supermassive object coincides with the bright radio/millimeter source
(Sgr~A*). The radiation is most probably of synchrotron origin and is believed
to be emitted by a hot, magnetized plasma interacting with the black hole. In
the radio band, the size of the source decreases towards shorter wavelengths
(\citealt{bower:2004}, \citealt{bower:2014}). At millimeter and shorter wavelengths, the
radiation is believed to originate in the immediate vicinity of the black hole.

Millimeter Very Long Baseline Interferometry (VLBI) observations
measured the average size of Sgr~A* to be $FWHM_{\lambda=1.3 {\rm mm}}=37^{+16}_{-10}
{\rm \mu as}$ \citep{doeleman:2008}. This size is comparable to the expected
diameter of the black hole "shadow" \citep{falcke:2000a}.  The detection of
the black hole shadow would be the first direct evidence of the black hole
in the Galactic center. Accurate measurements will be possible in the very near
future with enhanced high frequency VLBI observations involving an increasing
number of baselines. The detectability of the black
hole shadow at millimeter wavelengths will strongly depend on the source
geometry close to the black hole horizon.

At $\lambda>1$mm, the source structure is washed out due to scattering of
radio waves by electrons in the Milky Way \citep{bower:2014}.  Hence,
there is still debate concerning the exact origin of the
radio emission. Current theories that describe the sources’ electromagnetic
spectrum reasonably well may be divided into two subgroups: hot magnetized
accretion disks; and compact, magnetized jets (e.g. \citealt{narayan:1995}, 
\citealt{falcke:1993}, \citealt{yuan:2002}). Recently, general relativistic
magnetohydrodynamic (GRMHD) black hole accretion flow simulations, which often
produce relativistic jets (e.g. \citealt{mckinney:2006}, \citealt{noble:2007},
\citealt{beckwith:2008}, \citealt{sasha:2011} and many others), 
have become available and allow one to merge the two theories in a natural manner.

In \citet{moscibrodzka:2013}, we have presented such a merged GRMHD disk plus
jet model that is consistent with spectrum and sizes of Sgr~A*.  We have
demonstrated that jets produced by the GRMHD simulations produce the
characteristic flat radio spectrum, observed not only in Sgr A* but also in
jets in Active Galactic Nuclei, when the electrons in the jet are thermal with
a constant temperature and electrons in the disk have a constant {\em ratio} of
ion to electron temperature. The electron
temperature $T_{\rm e}$ is not explicitly modeled in GRMHD models and can be
treated as a model free parameter.
Nonetheless, the adopted prescription for
$T_{\rm e}$ is motivated by different physics in the disk and jet regions. The
accretion disk is more turbulent and less magnetized in comparison to the
jet. The heating of particles might be stronger in a strongly magnetized
plasma (e.g. \citealt{quataert:1998}, \citealt{quataert:1999}).

The present article is an extention of \citet{moscibrodzka:2013}. 
Here, we calculate spectra and images of the two
component (disk-plus-jet) models scaled to the size of the Sgr~A* system 
  in three dimensions.  Our
goal is to predict the appearance of the black hole ``shadow'' at
$\lambda$=1.3 mm, and to show that visible jets can be produced by GRMHD
simulations, with appropriate assumptions about electron thermodynamics.
Here, for the first time, we present millimeter
($\lambda$=1.3 mm) and radio ($\lambda$=3.5-13 mm) images of the merged model.

The present models are improved in comparison to our earlier study in three
respects.  1) We extend the previous axi-symmetric models to three
dimensions. 2) We investigate spectra and images produced by models
with various combinations of jet and disk electron temperatures and
for the first time present radio images of the jet produced by the
RIAF. 3) We include Compton scattering in the radiative transfer
scheme to model the spectral energy distribution up to $\gamma-$ray
energies. As we will show, the electron temperatures are constrained
by the observed NIR and X-ray fluxes.

The present coupled disk plus jet models are still simplified in two respects,
at least.  First, we do not yet associate the jet electron temperatures, or
e-p coupling in the disk, with any particular physical process. The disk and
the jet region are described by free but physically reasonable parameters.
In our model, we identify inflow and outflow regions and simply ascribe two
different electron heating mechanisms, which creates a somewhat artificially
sharp boundary between the two. 

Second, we assume that the electrons have a purely thermal distribution. The
thermal plasma assumption reduces the degrees of freedom in radiative transfer
modeling and simplifies our analysis. Electron acceleration into a nonthermal
component is nearly inevitable, however. It is the simplest explanation for
the NIR flaring component and could explain the high frequency NIR-X-ray 
spectrum uncovered by multiwavelength campaigns. We will consider a nonthermal
component in a future publication.

The article is organized as follows. In Sect.~\ref{sec:model} we
describe the dynamical and radiative transfer techniques used to calculate
spectra and images of the plasma. We also introduce the definition of
the jet and the prescription for electron temperatures in the jet and disk
zones. The spectra and images produced by the model as a function of
the electron temperature are presented in Sect.~\ref{sec:results}. We summarize the
results in Sect.~\ref{sec:diss}.

\section{Model of an accretion disk with a jet}~\label{sec:model}

\begin{figure}
\includegraphics[width=0.5\textwidth]{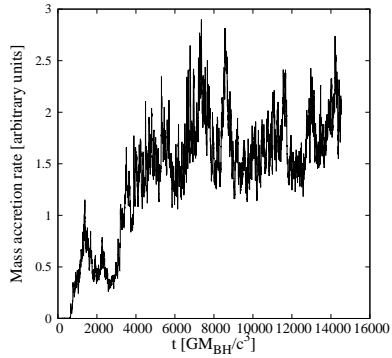}
\caption{Evolution of mass accretion rate (in arbitrary units) through the
  horizon during the entire simulation.}\label{fig:mdot}
\end{figure}

\subsection{Numerical model of accreting black hole}

As time-dependent model of a radiatively inefficient accretion flow
onto a black hole we use a three-dimensional (3-D) GRMHD simulation
carried out by Shiokawa 2013 (run b0-high in table 5.1 in Shiokawa
2013). The simulation started from a torus in hydrodynamic equilibrium
in the equatorial orbit around a rotating black hole \citep{fishbone:1976}. The
torus initially had a pressure maximum at $24 R_{\rm g}$ and an inner edge
at $12 R_{\rm g}$, where $R_{\rm g}=GM_{BH}/c^2$. It was seeded with a weak,
poloidal field that follows the isodensity contours (single loop
model, see Gammie et al. 2003). The dimensionless black hole spin was
$a_*\simeq 0.94$. The corresponding radius of the event horizon was
$r_h=1.348 R_{\rm g}$ and the inner-most stable circular orbit (ISCO) was located
at $r_{ISCO}=2.044 R_{\rm g}$.  The inner boundary of the computational
domain was just inside the event horizon and the outer boundary was at
$R_{out} = 240 R_{\rm g} \simeq 10.7$ AU, or an angular radius of $\simeq
1.2 \, {\rm mas}$ (assuming mass and distance of a black hole in the Galactic
center to be $M_{BH}=4.5\times10^6\msun$ and $D=8.5\times10^3$ PC,
respectively). The model was evolved for 14000 $R_{\rm g}/c$, which
corresponds to $87$h for the adopted black hole mass (for which the
time unit is $R_{\rm g}/c.=22.17s$). This time interval is equivalent to about
19 orbital periods of at $r=24 R_{\rm g}$.

\subsection{Jet and disk zones}\label{sec:zones}

\begin{figure}
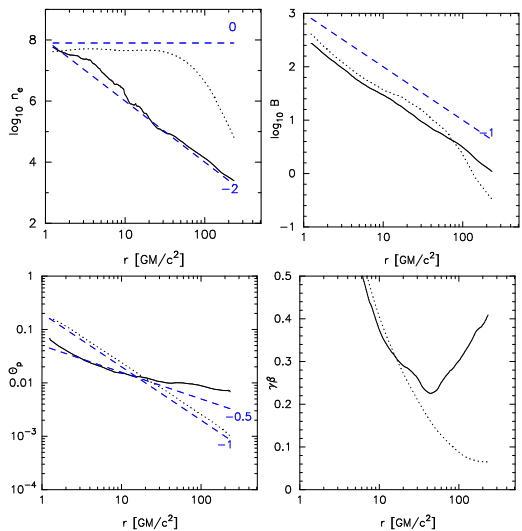

\includegraphics[width=0.25\textwidth,angle=-90]{f2a.ps}
\includegraphics[width=0.25\textwidth,angle=-90]{f2b.ps}\\
\includegraphics[width=0.25\textwidth,angle=-90]{f2c.ps}
\includegraphics[width=0.25\textwidth,angle=-90]{f2d.ps}
\caption{Shell- and time-averaged profiles of density, magnetic field
  strength, proton temperature ( defined as $\Theta_p=k_B T_p/m_p c^2$), 
and plasma speeds measured in the disk and in the jet zones.
The time averaging is done over later times during the simulation
($t=10000-10800\tu$). Solid and dotted lines show the averaged properties in the jet
sheath and the disk, respectively. The dashed lines indicate power-law
dependencies $r^p$, and are labelled with the parameter $p$.}\label{fig:prof}
\end{figure}

Fig.~\ref{fig:mdot} displays the mass accretion rate evolution over the entire
simulation. Early evolution of the magnetized torus is governed by the growth
of the magnetorotational instability which leads to the development of turbulence
and the accretion of plasma into the central object. At $t\approx10^4\tu$, the
flow has already reached a quasi-stationary state in which the disk is
accreting and well developed jets are present below and above the black hole
poles. We analyze a relatively short time sequence of the relaxed portion of
the simulation data
between $t=10000$ and $t=10800\tu$ ($\sim 5$ hours).

Following \citet{moscibrodzka:2013}, we formally define the jet zones
using the Bernoulli parameter $-hu_{\rm t}$, where $h=(\rho_0+e+P)/\rho_0$
is the fluid specific enthalpy and $u_{\rm t}$ is the covariant time
component of the four-velocity. We chose $-hu_{\rm t}$=1.02 to be a boundary
between jet and disk zones. For this value of the Bernoulli parameter the
jet narrows to a cone with a half opening angle of approximately $25
\degr$ and the definition of the jet excludes the unbound, subrelativistic
outflow produced by the outer regions of the disk. Regions with
$-hu_{\rm t} < 1.02$ are defined as an accretion flow/wind. One could also
adopt another description of jet zones (e.g. \citealt{dexter:2012},
\citealt{sadowski:2013a}). 
We argue that our definition of jet zones
is simple and effective and roughly separates outflow and inflow.

Fig.~\ref{fig:prof} shows the radial shell- and time-averaged profiles
(computed using Eq.~1 from \citealt{moscibrodzka:2013}) of density,
magnetic field strength, plasma temperature (we plot proton temperature 
in dimensionless units $\Theta_p \equiv k T_p/(m_p c^2)$) where
$\Theta_e \equiv k T_e/(m_e c^2)$, computed assuming that $T_{\rm e}=T_{\rm p}$,and 
$T_{\rm p}$ is computed directly from the simulation) and plasma speed in the
jet sheath (i.e. omitting the strongly magnetized but nearly empty jet
spine) and in the accretion flow/wind.
 
In the disk zones, the density and magnetic field follow $r^{0}$ and 
$r^{-1}$ dependencies, respectively. The radial 
profile of the density is rather shallow but it is consistent
with inflow/outflow RIAF models in which the density profile usually
follows a $r^{-3/2+s}$ law where s is a parameter. 
In time-dependent MHD models parameter $s$ typically ranges from 0.4 to 1 
(\citealt{begelman:2012}, \citealt{narayan:2012}, \citealt{yuan:2014}) 
and one would expect $n_e \sim r^{-0.5}$. In our model, the density 
has an even shallower profile due to a formal definition 
of the disk and jet region and due to the fact that
our initial disk is rather small from the beginning (initial conditions
artifact). 

In the jet zones, the density
and magnetic field radial distribution roughly follow $r^{-2}$ and
$r^{-1}$ dependencies, respectively. The magnetic field strength
radial dependence in the 3-D model (but not in the 2-D models) 
reproduces the semi-analytical models of jets needed to explain flat-spectrum
radio cored (e.g. \citealt{bruyn:1976}, \citealt{blandford:1979}, 
 \citealt{falcke:1995}, \citealt{falcke:1996}).
The plasma outflowing in the jet sheath starts to
accelerate at $r\sim50 R_{\rm g}$ and reaches speeds of
$\gamma\beta \approx 0.4$ (where $\beta$ is speed measured 
by the normal observer) at the outer boundary of the simulation.
The accretion flow temperatures are virial ($T_{\rm p} \sim
r^{-1}$) and in the jet $T_{\rm p} \sim r^{-0.5}$. 
Similar dependencies
were found in our axisymmetric models (Fig.~1 in
\citealt{moscibrodzka:2013}), which proves the robustness of the two
component, spine and sheath, jet solution.

\subsection{Model free parameters and comparison of the model to observations}\label{param}

The radiative properties of the dynamical model are studied by combining the
magnetohydrodynamic model with post-processing radiative transfer
computations. Calculations are carried out using the same tools as in
\citet{moscibrodzka:2009}. The spectral energy distributions (SEDs) and images
of the plasma are computed using a Monte Carlo code for relativistic radiative
transfer \citep{dolence:2009} and a ray tracing radiative transfer scheme
\citep{noble:2007}. The radiative models include synchrotron emission,
absorption and inverse-Compton scattering. Synchrotron emissivities and absorptivities along
geodesics trajectories are computed assuming that radiating electrons have a
purely thermal energy distribution function. Accurate synchrotron emissivity
functions for the thermal electrons are taken from \citet{leung:2011}. 
The radiative transfer codes use
all variables, except $T_{\rm e}$, directly from the accretion flow model.

As in \citet{moscibrodzka:2013}, the following prescription for $T_{\rm e}$ is
adopted. In the disk zones (defined in Sect.~\ref{sec:zones}), the electron 
temperature is described using a standard assumption that the electron
temperature is a fraction of the proton temperature:
\begin{equation}\label{eq:1a}
\frac {k_b T_e}{m_e c^2} = \left(\frac{k_b T_p}{\mu m_p}\right) \left(\frac{m_p}{m_e}\right) \frac{1}{
  (T_p/T_e)_{disk}},
\end{equation}
where $T_p$ is taken from the simulation.
In the jet zones, we assume a constant electron temperature
\begin{equation}\label{eq:1b}
\frac{k_b T_e}{m_e c^2} = \Theta_{e,j}  = const.
\end{equation}
The disk $\tratd$ (Eq.~\ref{eq:1a}) and jet $\tj$ (Eq.~\ref{eq:1b}) 
are the model-free parameters.
In our model, we identify disk and jet regions and simply ascribe two
different electron heating mechanism, which creates an
artificially sharp boundary, so the electron temperature does not transition
smoothly from one region to another.
The fixed model parameters are the black hole mass, 
and the distance to the source. The mass of
the black hole sets the length scale of the system $\rg$, 
and the distance
allows one to translate the absolute system luminosity to an observed luminosity
and the physical size of the system to the angular size on the sky.
We scale the models to the Galactic center supermassive black hole. We adopt
$M_{\rm BH}=4.5\times10^6 \msun$, and $D=8.5\times 10^3$ PC.
We then vary three free model parameters: the constant electron
temperature in the jet ($\tj$), the electron to proton temperature
ratio in the disk ($\tratd$), and the inclination of the observer with
respect to the black hole spin ($i$).  All models are normalized to
produce the 2.4 Jy flux at $\lambda=$1.3 mm \citep{doeleman:2008} and
the normalization is done by changing the mass accretion rate $\mdot$
(i.e. multiplying the matter densities by a constant). 
Notice that in case of Sgr~A* the range of possible $\tj$ and $\tratd$
is constrained because $\dot{M}$ is constrained by observations of Faraday
rotation of the polarized radio emission
($2\times10^{-9}<\dot{M}<2\times10^{-7} \, \mdotu$, \citealt{bower:2005}, 
\citealt{marrone:2007}). Finally, when modeling
images of the simulations, we usually assume that the position angle PA of the black hole
spin axis projected on the sky plane is $\xi=0\degr$ East of North direction
($\xi=0$ corresponds to the black hole spin axis inclined at $PA\sim 60\degr$ 
relative to the Galaxy's rotation axis, \citealt{li:2013}).

The scheme to scale the model to the real object is almost identical to the one presented
in \citet{moscibrodzka:2009} and \citep{noble:2007} with a few exceptions: the
jet electron temperature is now an independent free parameter; we measure the 
sizes of the model for $\lambda \ge 1.3$ mm; we limit our model survey to only one
black hole spin (the spin survey in 3-D is computationally expensive and
beyond the scope of the present work).

In Sects.~\ref{sub:1}, \ref{sub:2},
and~\ref{sub:3} we show general dependencies of the model appearance and
spectra on the parameters based on a few chosen examples. In
Sect.~\ref{sub:4}, we show a model that best describes emission and size of 
Sgr~A*.

\section{Results}\label{sec:results}

We have studied 45 radiative transfer models with various combinations of
$\tj$ and $\tratd$ observed at inclination angles of
$i=30\degr,60\degr,90\degr$. All model parameters together with "normalizing"
accretion rates are listed in Table~\ref{tab:1}. 

\begin{table}[h!]
\centering
\caption{List of radiative transfer model parameters.}\label{tab:1}
\begin{tabular}{c c c c c }
\hline\hline
model \# & $i$ & $\tj$ & $(\frac{T_{\rm p}}{T_{\rm
    e}})_{\rm disk}$ &$\mdot [\mdotu]$ \\
\hline
1&&          & $5$  & $4.0\times 10^{-9}$ \\
2&&          & $10$ & $1.3\times 10^{-8}$ \\
3&$90\degr$&   $10$ & $15$ & $3.7\times 10^{-8}$ \\
4&&          & $20$ & $7.2\times 10^{-8}$ \\
5&&          & $25$ & $1.0\times 10^{-7}$ \\
\hline
6&&          & $5$  & $3.9\times 10^{-9}$ \\
7&&          & $10$ & $1.2\times 10^{-8}$ \\
8&$90\degr$&   $20$ & $15$ & $2.9\times 10^{-8}$ \\
9&&          & $20$ & $4.6\times 10^{-8}$ \\
10&&         & $25$ & $5.7\times 10^{-8}$ \\
\hline
11&&          & $5$  & $3.9\times 10^{-9}$ \\
12&&          & $10$ & $1.1\times 10^{-8}$ \\
13&$90\degr$&   $30$ & $15$ & $2.4\times 10^{-8}$ \\
14&&          & $20$ & $3.6\times 10^{-8}$ \\
15&&          & $25$ & $4.3\times 10^{-8}$ \\
\hline
\hline
16&&          & $5$  & $3.9\times 10^{-9}$ \\
17&&          & $10$ & $1.3\times 10^{-8}$ \\
18&$60\degr$&   $10$ & $15$ & $3.2\times 10^{-8}$ \\
19&&          & $20$ & $6.1\times 10^{-8}$ \\
20&&          & $25$ & $9.0\times 10^{-8}$ \\
\hline
21&   &       & $5$  & $4.4\times 10^{-9}$ \\
22&   &       & $10$ & $1.2\times 10^{-8}$ \\
23&$60\degr$&   $20$ & $15$ & $2.7\times 10^{-8}$ \\
24&  &        & $20$ & $4.2\times 10^{-8}$ \\
25&  &        & $25$ & $5.4\times 10^{-8}$ \\
\hline
26&&          & $5$  & $4.3\times 10^{-9}$\\
27&&          & $10$ & $1.2\times 10^{-8}$\\
28&$60\degr$&   $30$ & $15$ & $2.4\times 10^{-8}$\\
29&&          & $20$ & $3.5\times 10^{-8}$\\
30&&          & $25$ & $4.2\times 10^{-8}$\\
\hline
\hline
31&   &       & $5$  & $5.6\times 10^{-9}$\\
32&   &       & $10$ & $1.4\times 10^{-8}$\\
33&$30\degr$&   $10$ & $15$ & $1.0\times 10^{-8}$\\
34&  &        & $20$ & $5.7\times 10^{-8}$\\
35&  &        & $25$ & $8.4\times 10^{-8}$\\
\hline
36&&          & $5$  & $5.6\times 10^{-9}$\\
37&&          & $10$ & $1.37\times 10^{-8}$\\
38&$30\degr$&   $20$ & $15$ & $2.7\times 10^{-8}$\\
39&&          & $20$ & $4.1\times 10^{-8}$\\
40&&          & $25$ & $5.2\times 10^{-8}$\\
\hline
41&  &        & $5$  & $5.6\times 10^{-9}$\\
42&  &        & $10$ & $1.3\times 10^{-8}$\\
43&$30\degr$&   $30$ & $15$ & $2.4\times 10^{-8}$\\
44&  &        & $20$ & $3.4\times 10^{-8}$\\
45&  &        & $25$ & $4.1\times 10^{-8}$\\
\hline
\end{tabular}
\end{table}

\subsection{Disk temperature $\tratd$ dependence}\label{sub:1}

\begin{figure}
\includegraphics[width=0.16\textwidth]{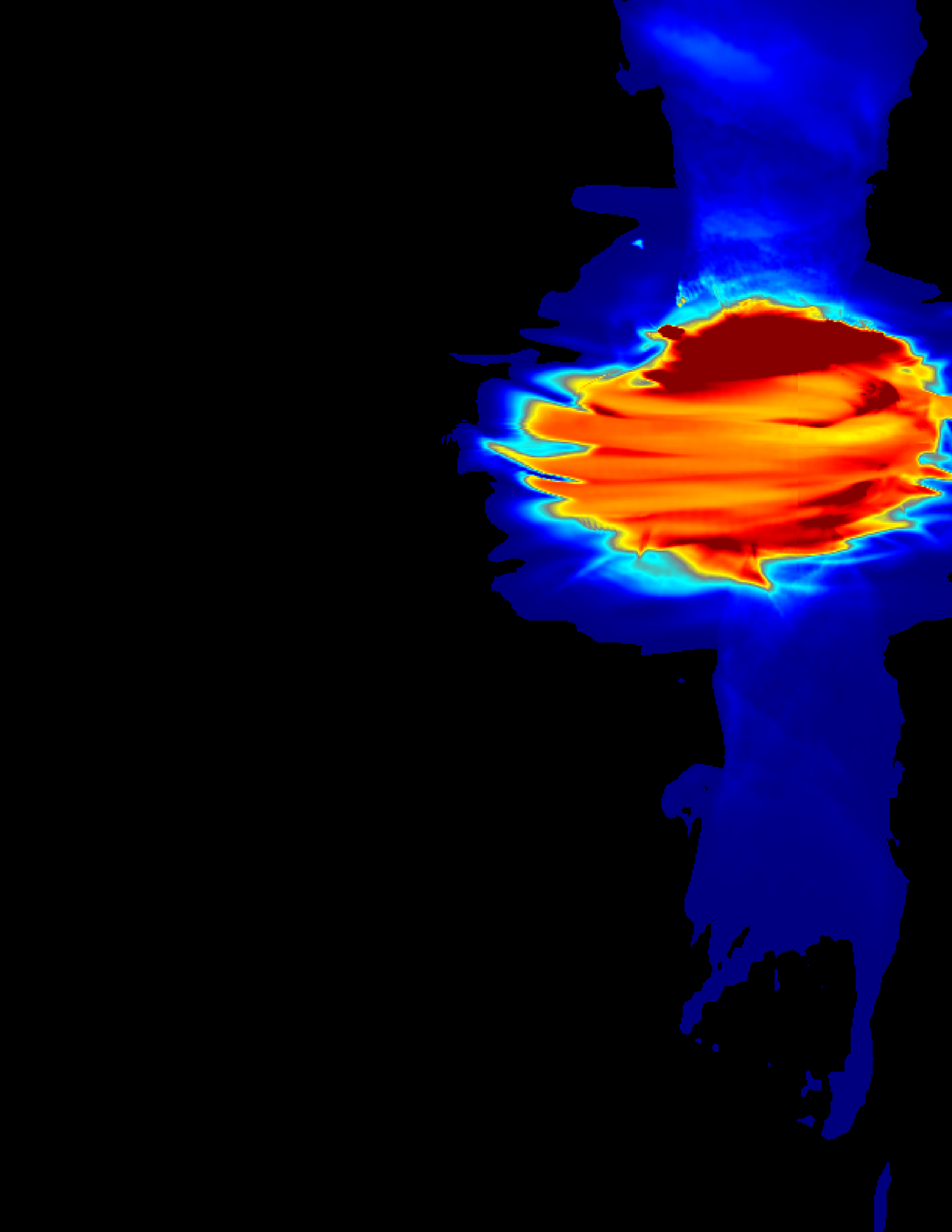}
\includegraphics[width=0.16\textwidth]{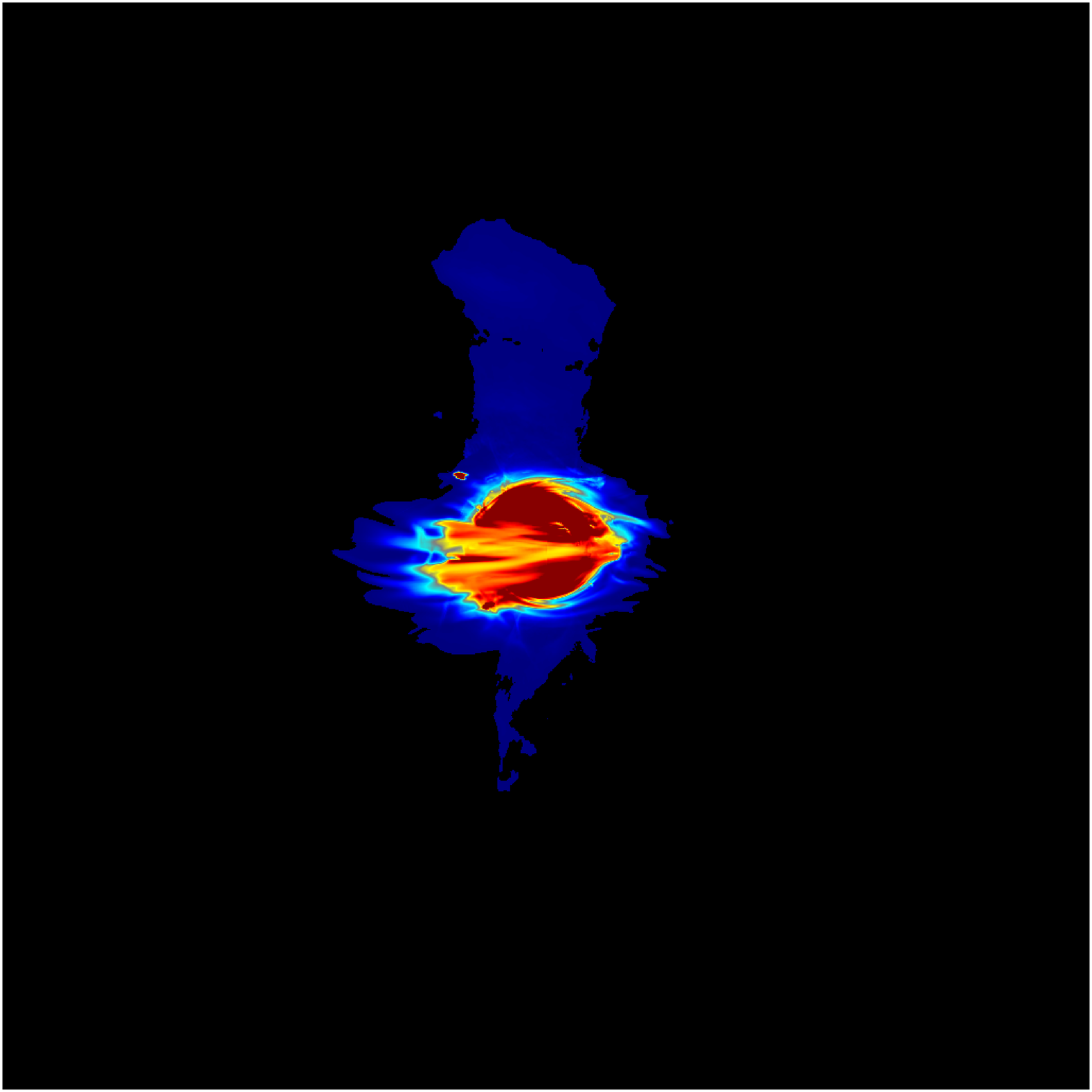}
\includegraphics[width=0.16\textwidth]{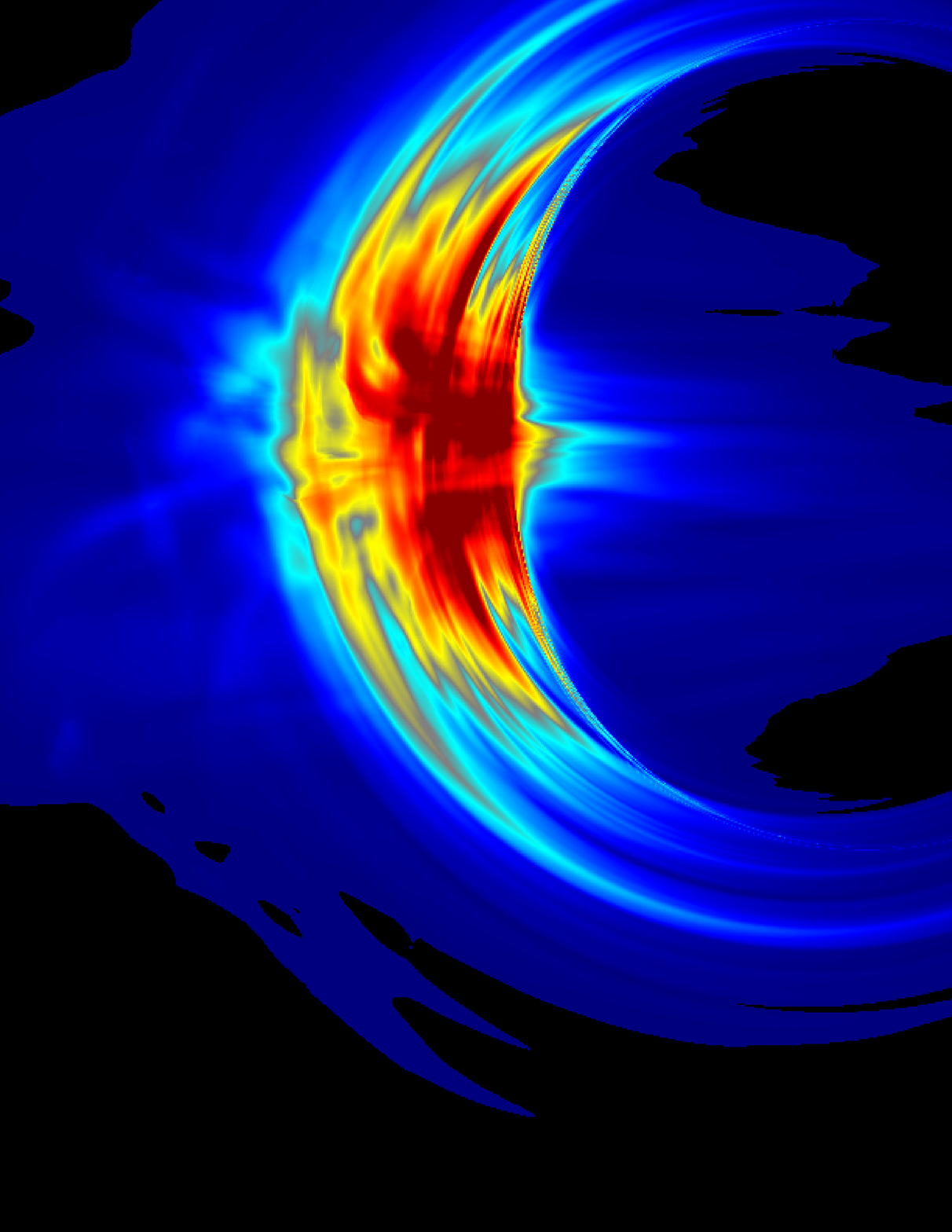}\\
\includegraphics[width=0.16\textwidth]{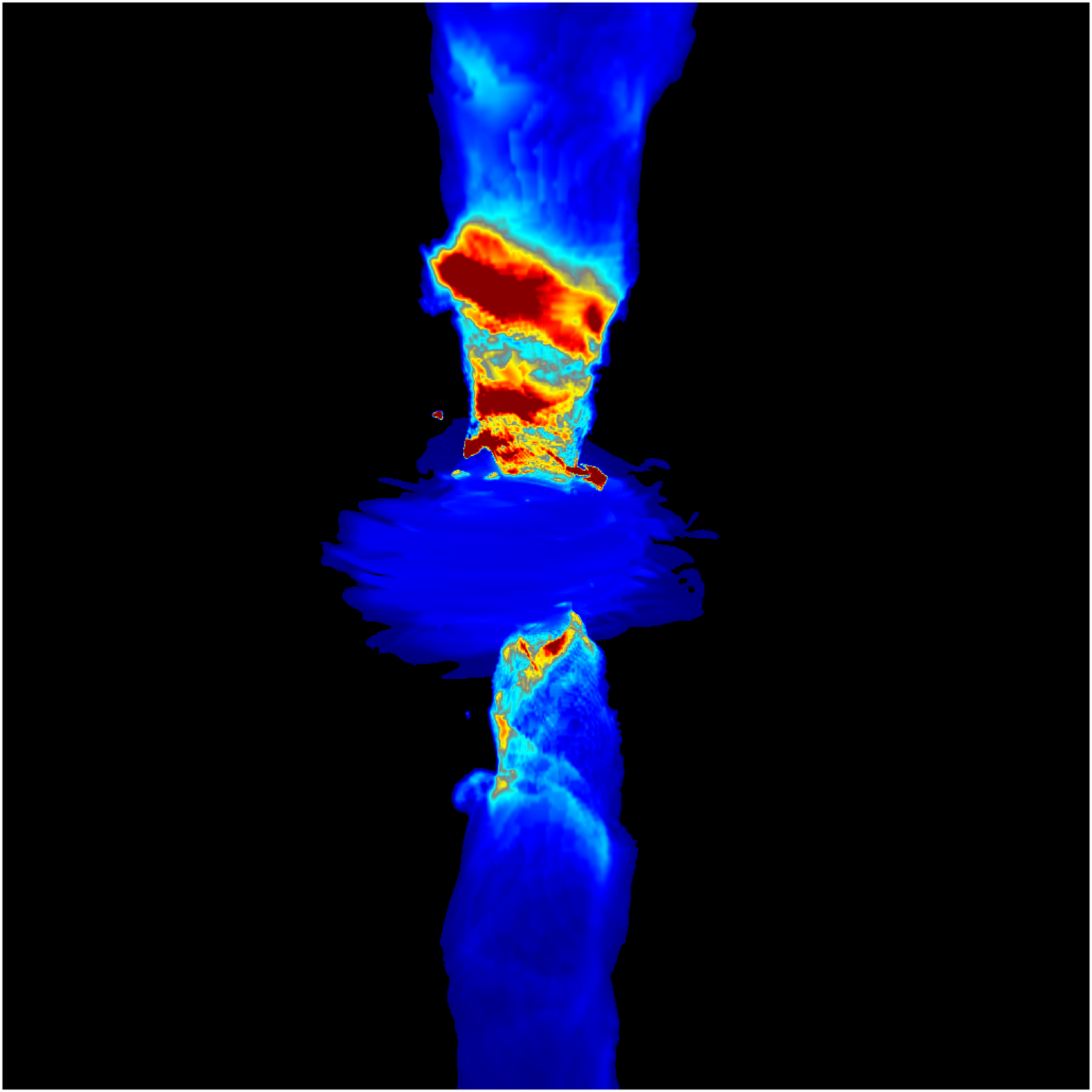}
\includegraphics[width=0.16\textwidth]{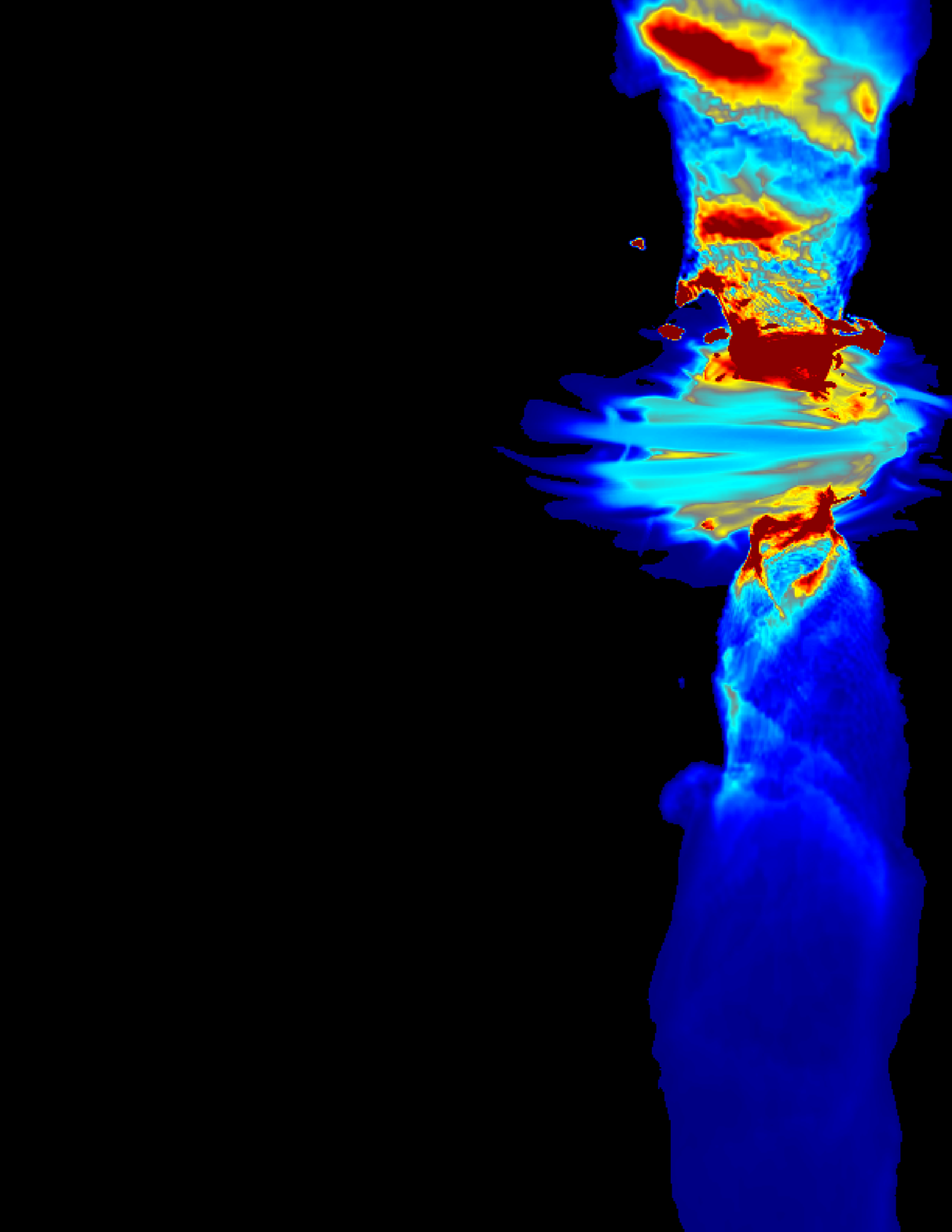}
\includegraphics[width=0.16\textwidth]{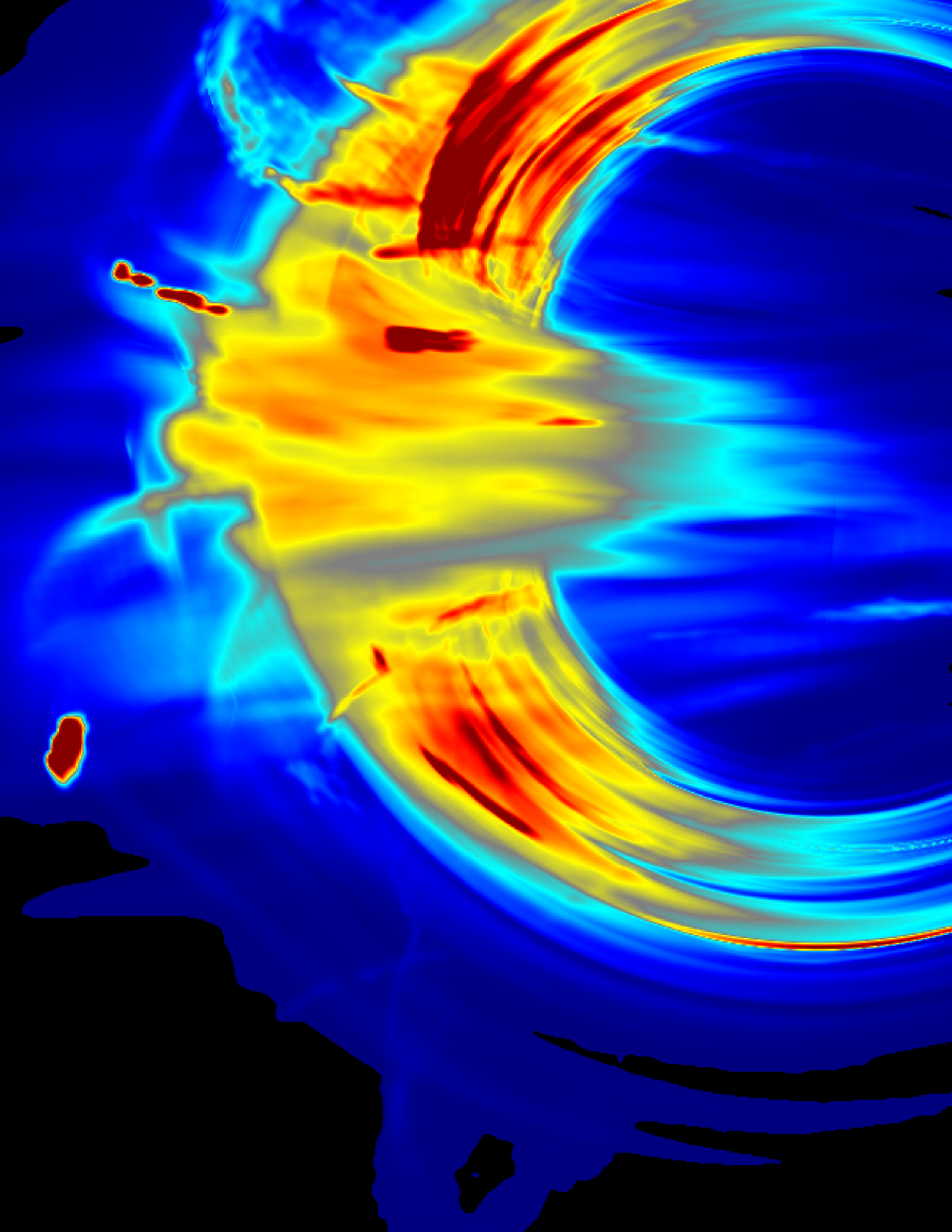}\\
\includegraphics[width=0.16\textwidth]{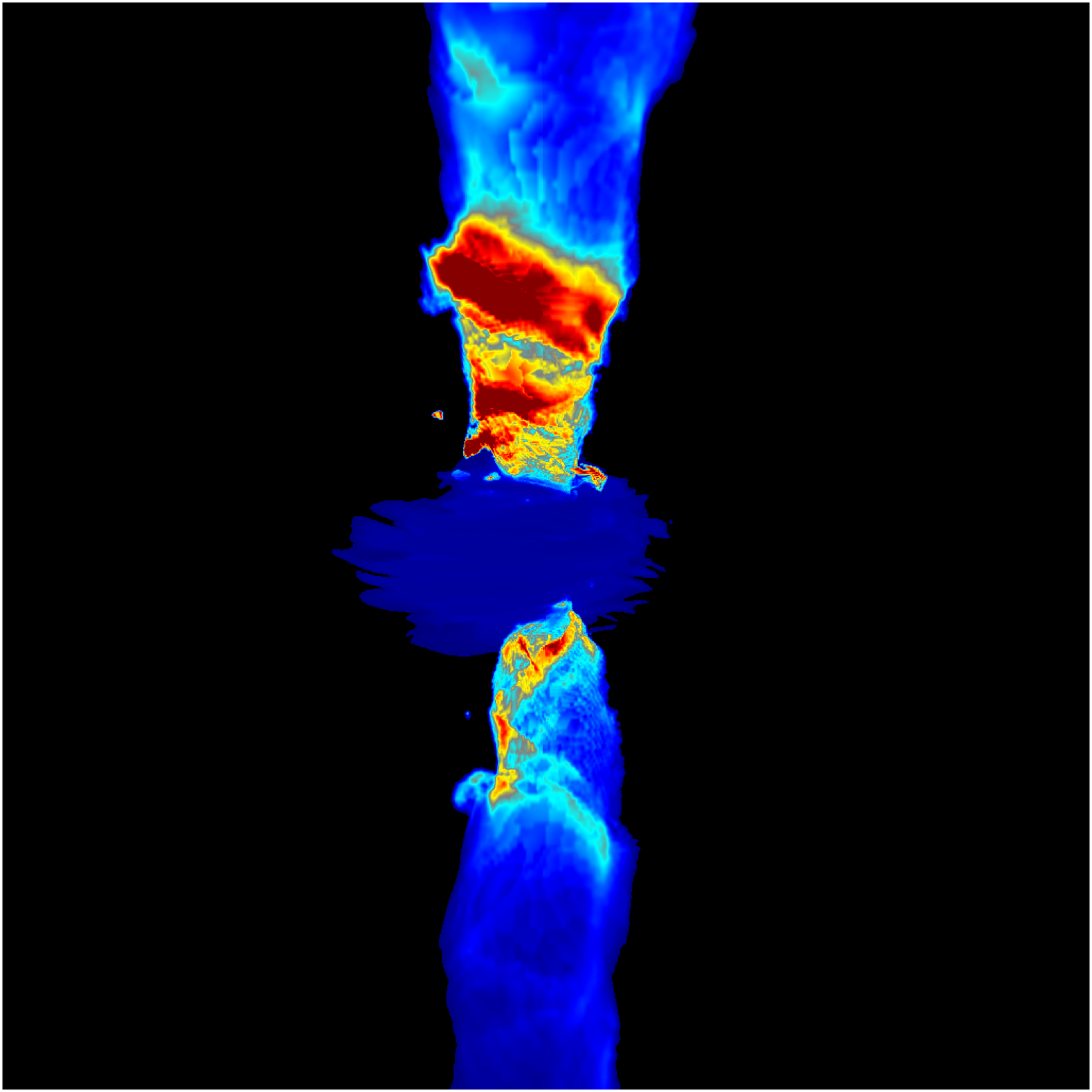}
\includegraphics[width=0.16\textwidth]{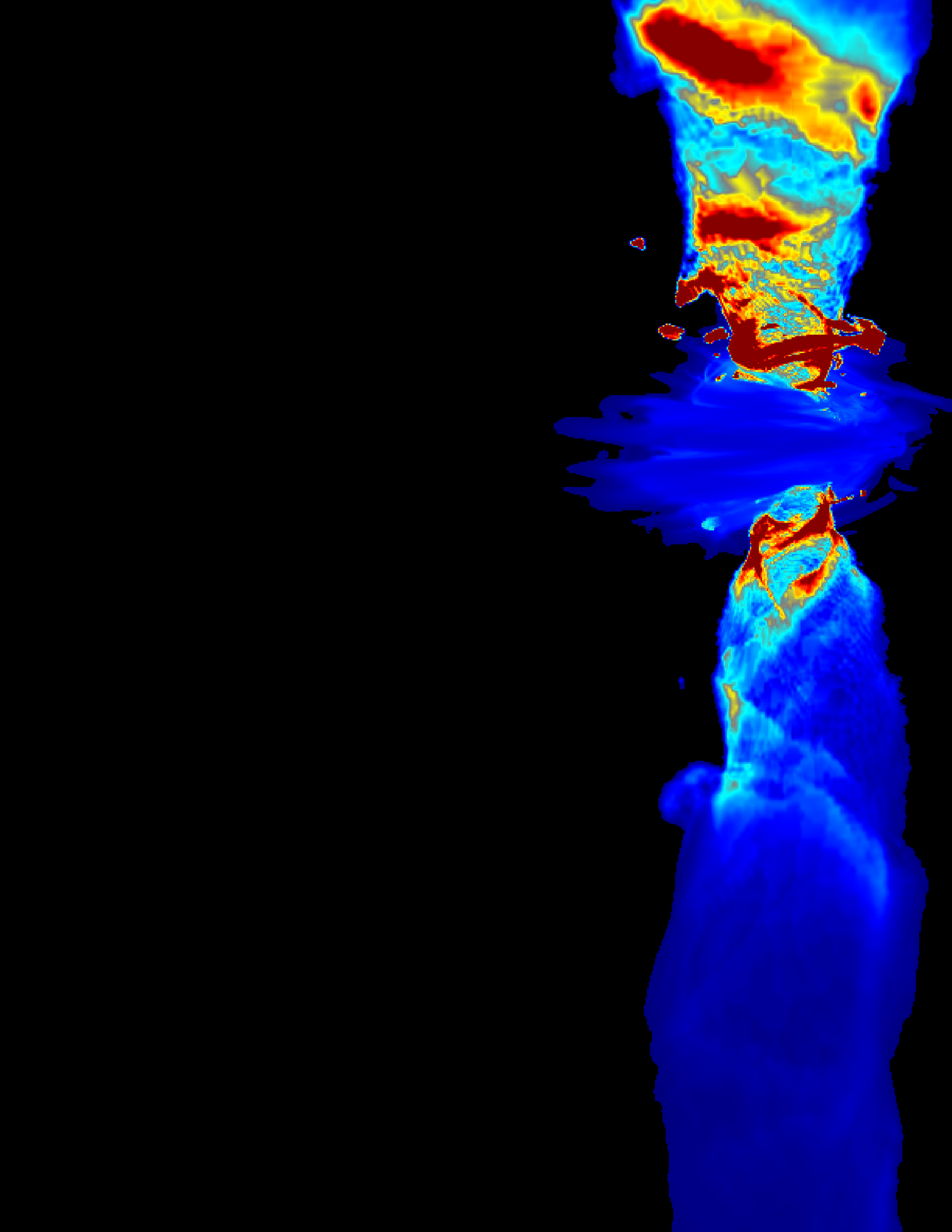}
\includegraphics[width=0.16\textwidth]{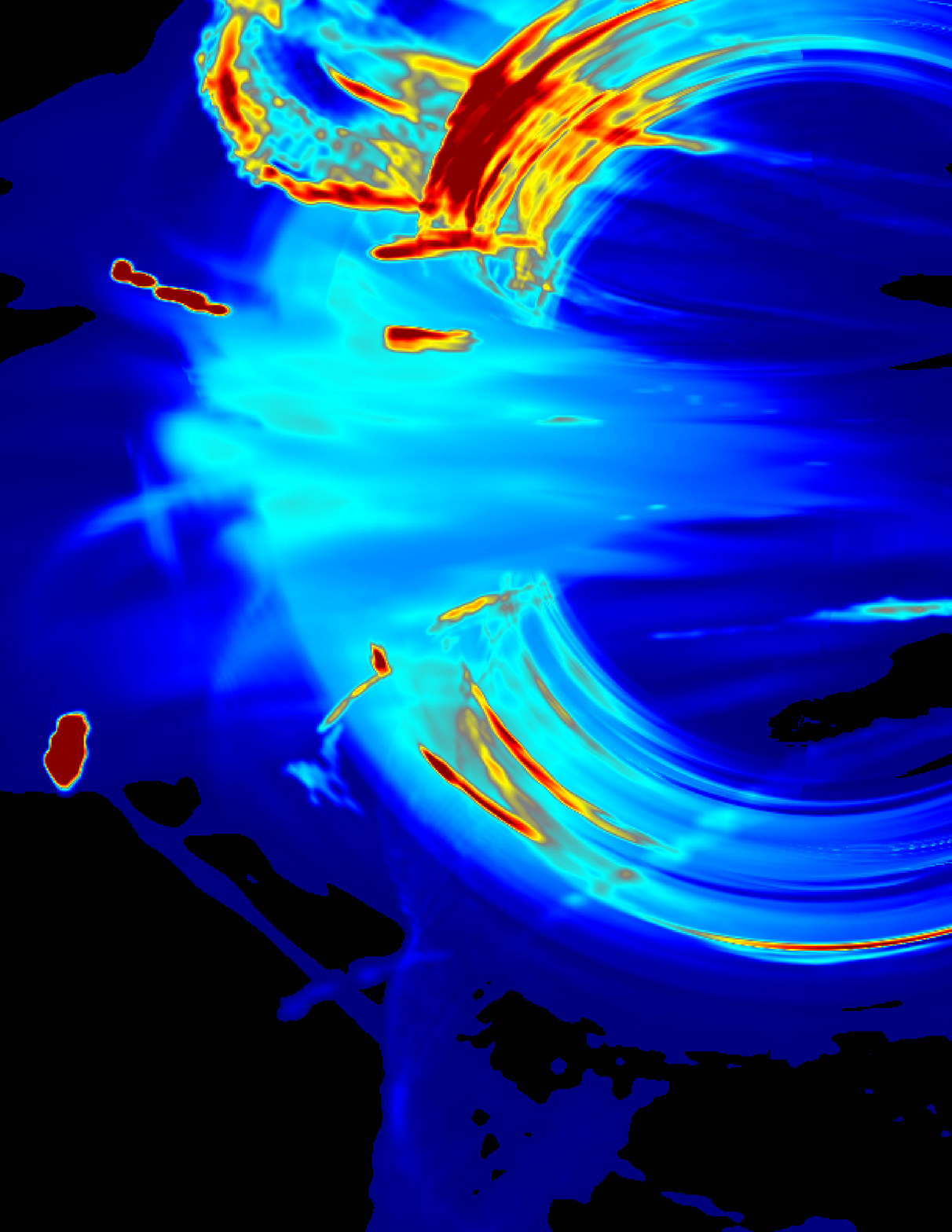}\\
\includegraphics[width=0.49\textwidth]{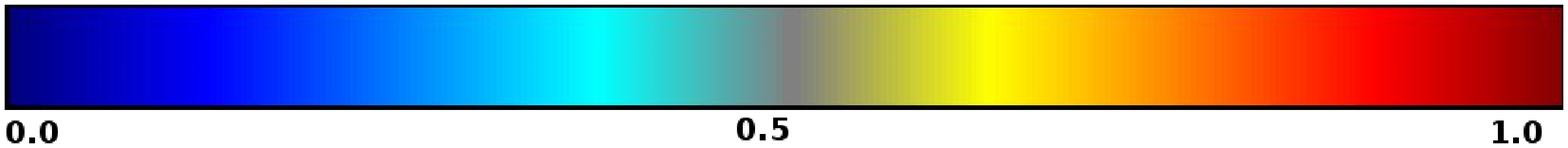}\\
\includegraphics[width=0.45\textwidth,angle=-90]{f3k.eps}
\caption{Dependency on $\tratd$. Upper part: Panels from top to bottom
  display "infinite"-resolution images of models with
  ($\tratd,\tj$)=(5,20), (15,20), and (25,20) at a viewing angle
  $i=90\degr$. Left, middle, and right columns show the model
  appearance at $\lambda=$13, 7 and 1.3 mm, respectively. Colors code
  the radiation intensity on a linear scale. The field of view of the
  left and middle panels is $200\times200 {\rm \rg}$ (approximately
  $1\times1 {\rm mas}$), and the right panel's field of view is
  $20\times20 {\rm \rg}$. Lower part: spectrum emitted by the
  various models. The normalization is marked with a black
  point.~\label{img_trat}}
\end{figure}

Fig.~\ref{img_trat} (upper part) shows "infinite"-resolution (i.e. not
accounting for interstellar scattering or finite instrumental resolution) 
images of models with
($\tratd$,$\tj$)=(5,20), (15,20) and (25,20) at $\lambda$=13, 7, 1.3 mm.  All
panels display the same single snapshot of the simulation. 
Fig.~\ref{img_trat} (lower part) shows spectra emitted by the three 
models presented here.
The mass accretion rate required to renormalize models to produce the same
flux $\lambda=$1.3 mm
(marked in the lower part of Fig.~\ref{img_trat} as a black point)
increases by a factor of 15 from model with ($\tratd$,$\tj$)=(5,20) to
($\tratd$,$\tj$)=(25,20) ($\mdot=3.9\times10^{-9} - 5.7\times10^{-8} \mdotu$).

The appearance of the model strongly depends on the assumed electron
temperatures, as expected.  In general, in all models the size of the emission
region increases with wavelength but the rate of increase with wavelength is
somewhat different. In models with ($\tratd$,$\tj$)=(5,20) the disk is brighter
than the jet at radio wavelengths.  The synchrotron photosphere at 7 and 13 mm
is compact and symmetric for the edge-on view.  On the other hand, in models
with ($\tratd$,$\tj$)=(25,20) the emission at radio wavelengths is primarily
produced by the jet plasma. The emission region in this case is patchy. 
Evidently, the jet photosphere is
significantly more extended in comparison to the disk photosphere.  The radio
spectra emitted by the model in which the jet emission dominates has a nearly
{flat-to-inverted} slope ($\alpha_\nu\sim0.3$, where $F_\nu\sim\nu^{\alpha}$) in
accordance with the analytical predictions; 
the spectral slope produced by the disk
photosphere is significantly steeper ($\alpha_{\nu}\sim1$).

At sub-millimeter wavelengths, the plasma becomes optically thin and the spectrum
of all models undergoes a slope inversion. At $\lambda=$1.3 mm, the emission
comes from the direct vicinity of the black hole horizon.  In the right panels
in Fig.~\ref{img_trat}, the shadow of
the black hole is visible in all models. At this point, the
general and special relativistic effects are equally important in shaping the
appearance of the emitting region.  The geometry of the emitting region
strongly depends on the model parameters. In models with
($\tratd$,$\tj$)=(5,20) the source angular size is smaller than the diameter
of the black hole shadow.
The emission is dominated by the left side of the disk where the
orbiting plasma is approaching the observer and the emission is
therefore strongly Doppler boosted.
The image of the plasma might be described by a
Gaussian or a crescent shape and the image morphology is similar to images
presented in \citet{noble:2007}, 
\citet{broderick:2009a}, \citet{dexter:2010},
\citet{roman:2012} and \citet{kamruddin:2013}.  
In models with ($\tratd$,$\tj$)=(25,20) the
black hole shadow becomes more evident because the plasma emission is more
extended. In this case the images are also more patchy, similar to emission at
$\lambda=$7 and 13 mm. The models are not well represented by
a Gaussian or a crescent.

Interestingly, in all models emission at energies higher than NIR is produced
primarily via first order inverse-Compton scattering.  We find that the
scattering occurs mainly in the accretion disk (due to its higher optical
depth) within a radial range $r=4-10 \rg$.  Therefore, the spectrum at high
energies strongly depends on the disk electron temperatures, i.e.~on
$\tratd$. The X-ray to mm flux ratio increases by more than an order of
magnitude from models with ($\tratd$,$\tj$)=(5,20) to
($\tratd$,$\tj$)=(25,20). These conclusions could change for models that
incorporate a nonthermal component in the electron distribution function.

\subsection{Jet temperature $\tj$ dependence}\label{sub:2}

\begin{figure}
\includegraphics[width=0.16\textwidth]{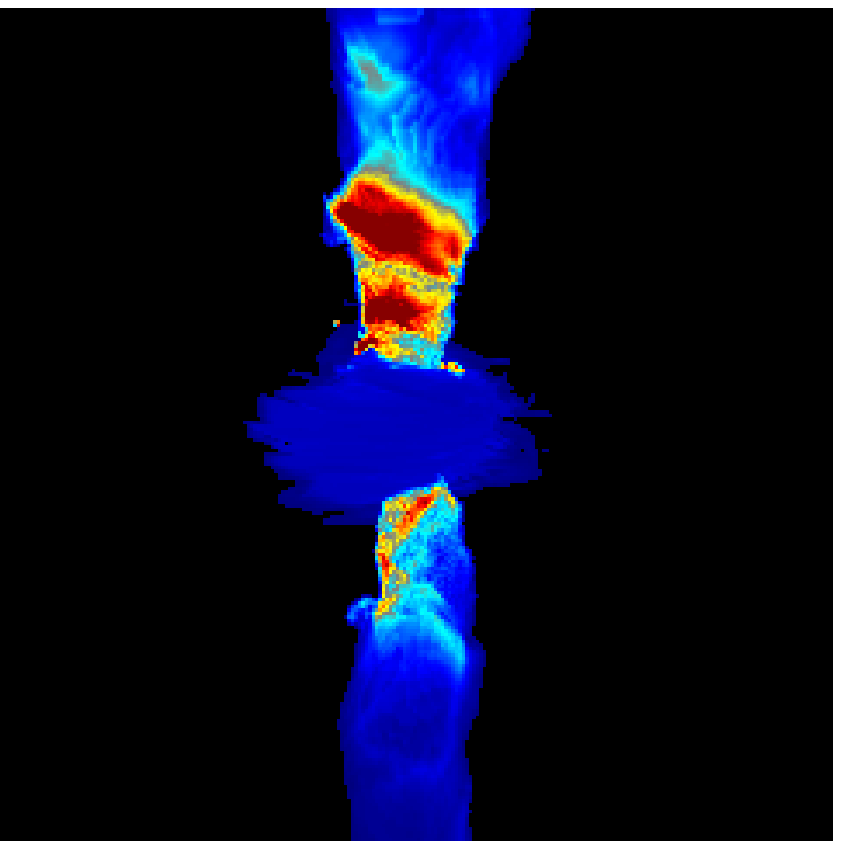}
\includegraphics[width=0.16\textwidth]{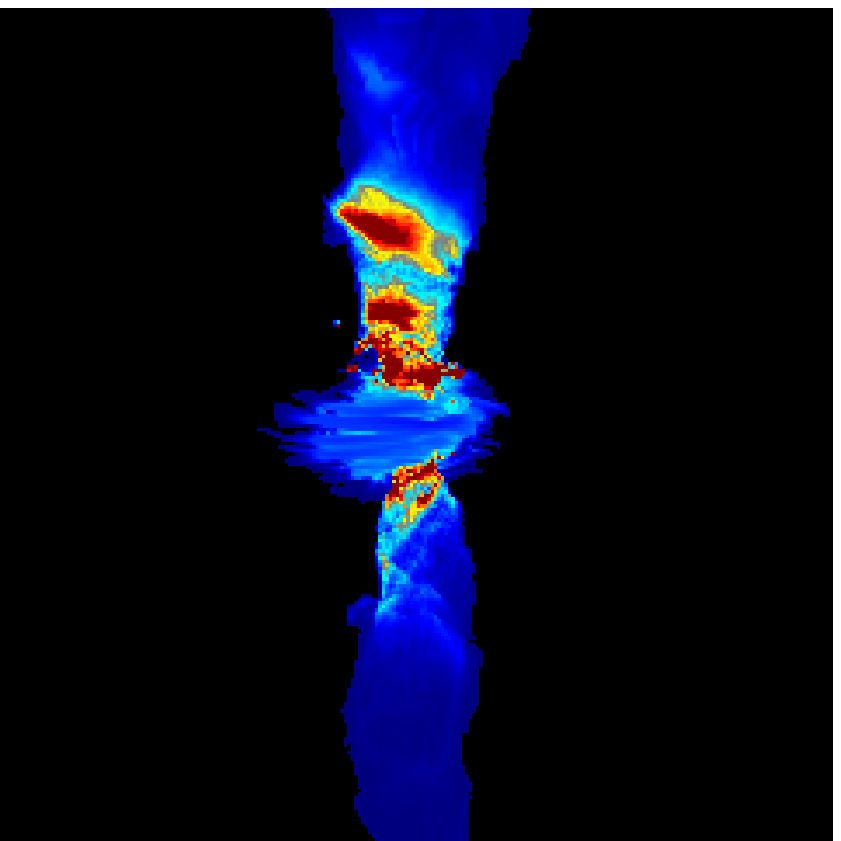}
\includegraphics[width=0.16\textwidth]{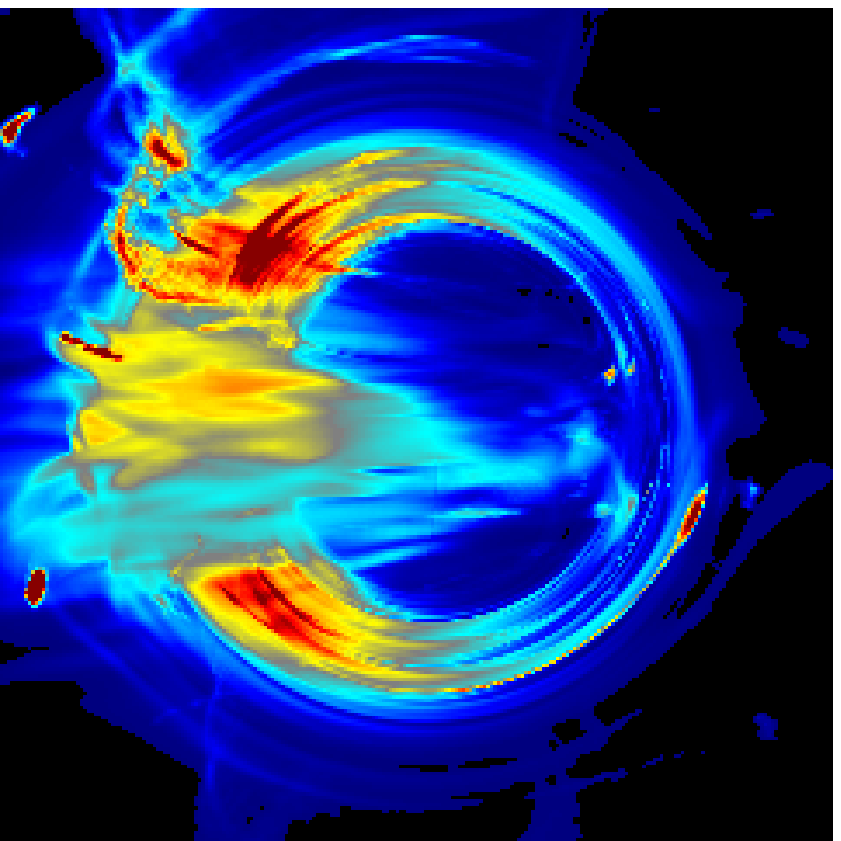}\\
\includegraphics[width=0.16\textwidth]{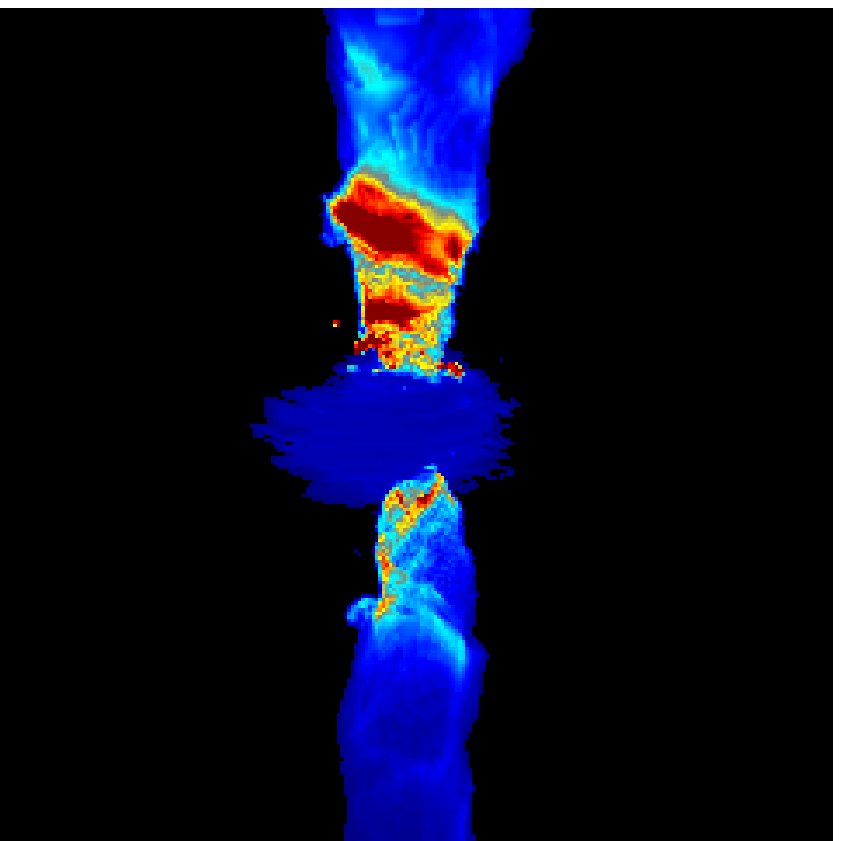}
\includegraphics[width=0.16\textwidth]{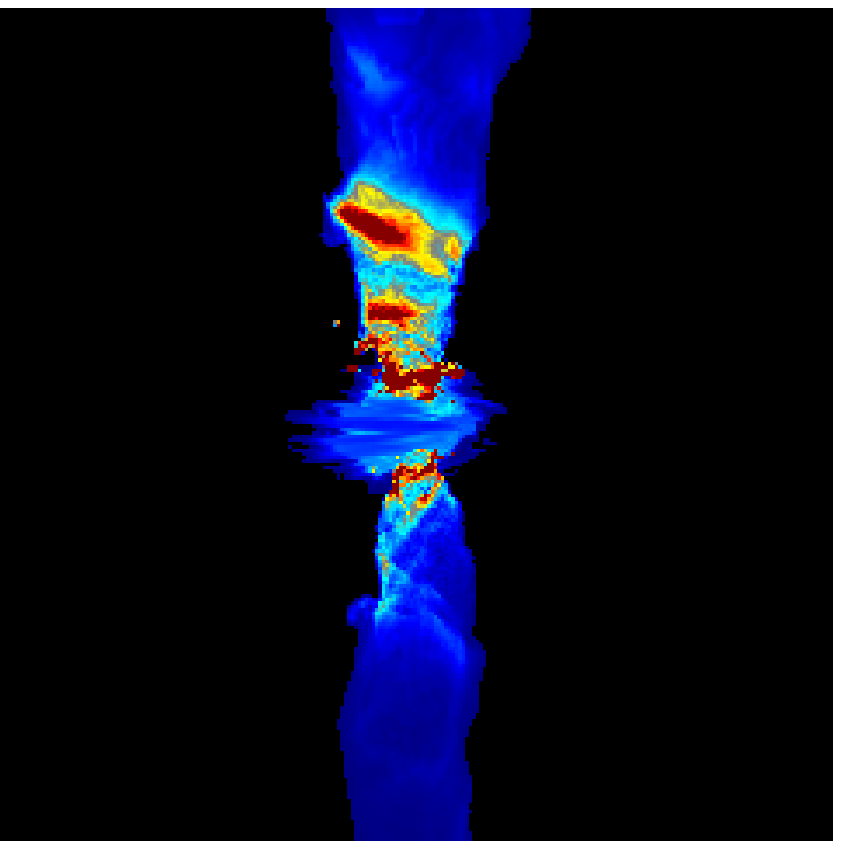}
\includegraphics[width=0.16\textwidth]{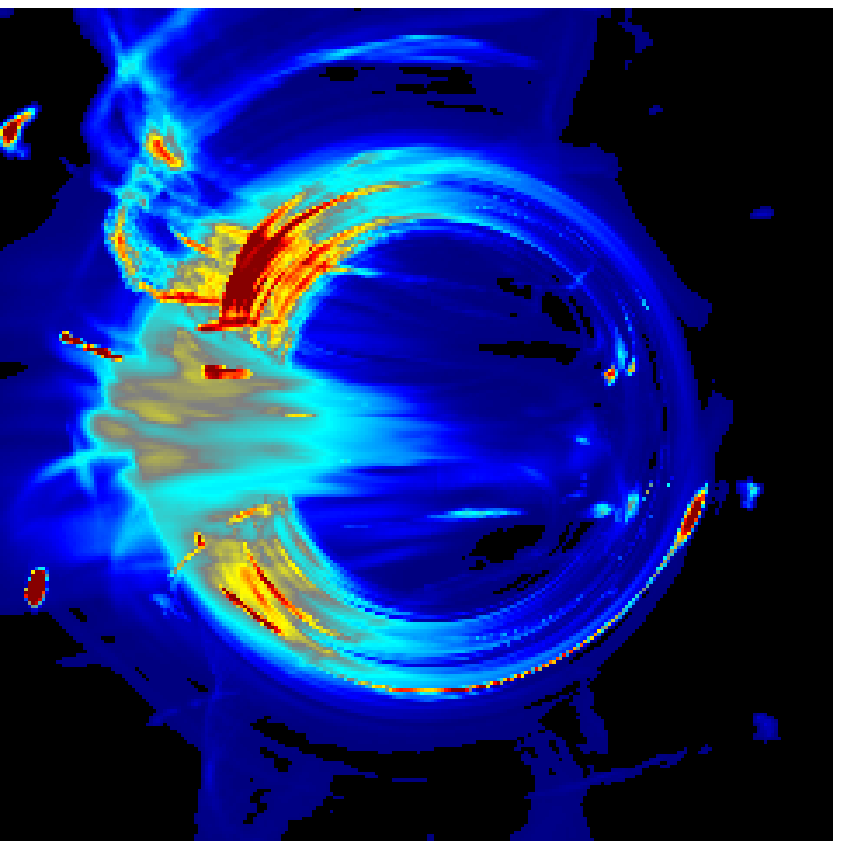}\\
\includegraphics[width=0.16\textwidth]{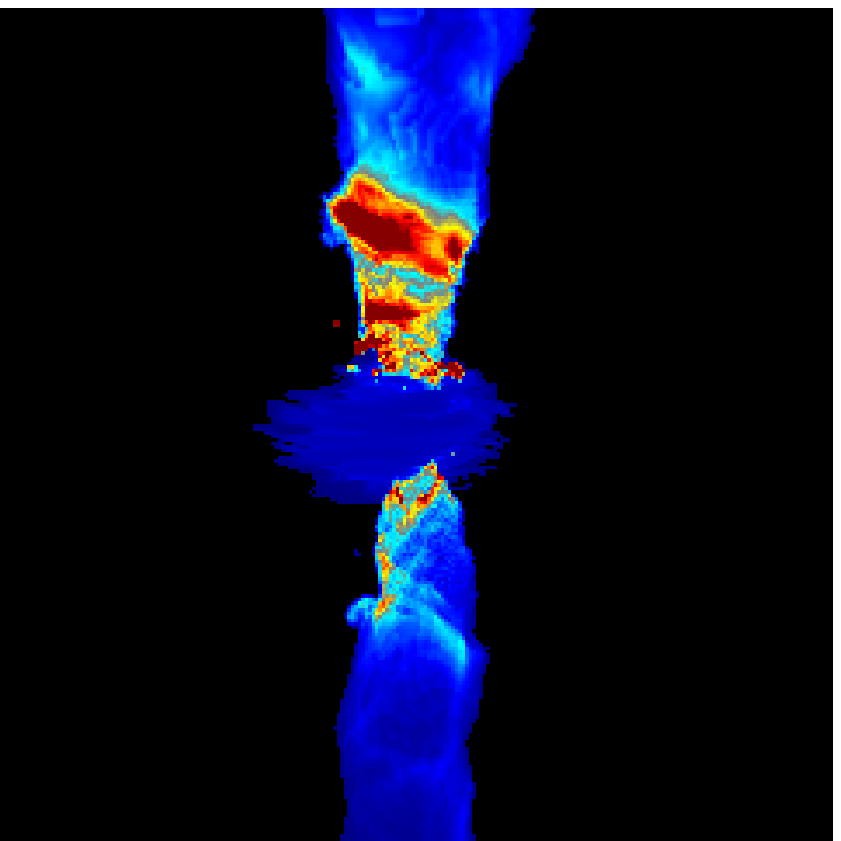}
\includegraphics[width=0.16\textwidth]{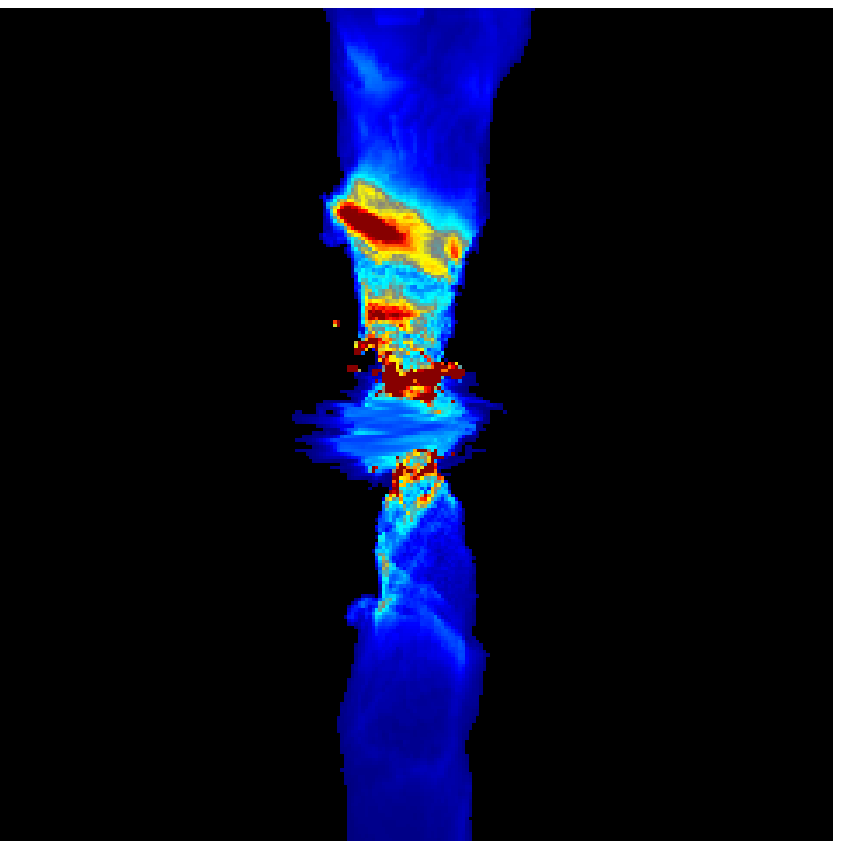}
\includegraphics[width=0.16\textwidth]{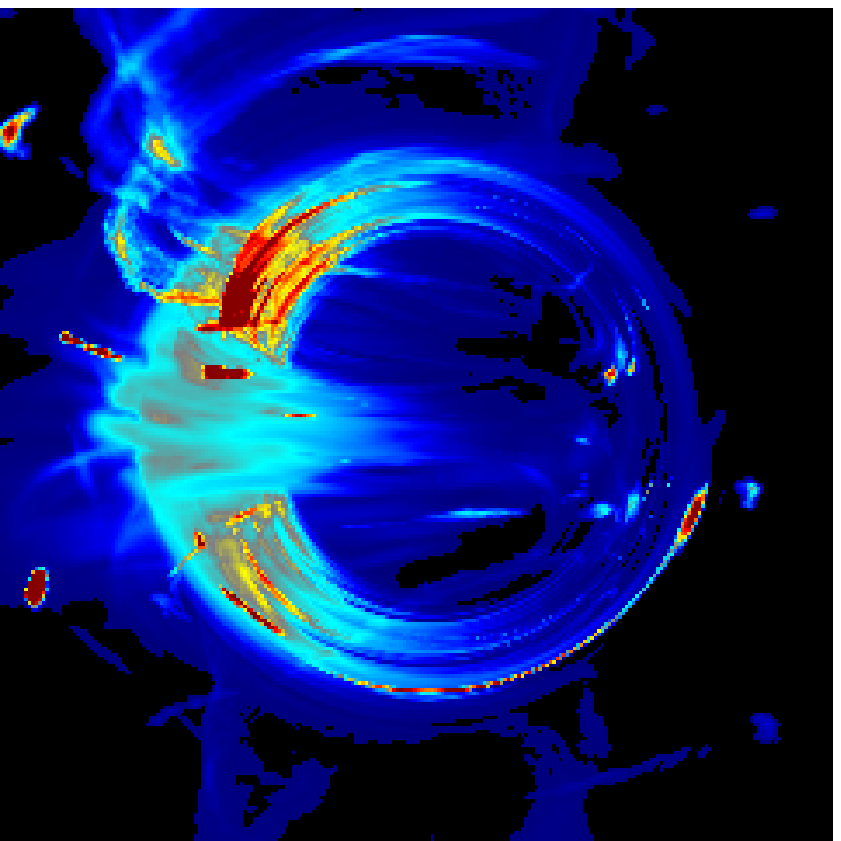}\\
\includegraphics[width=0.49\textwidth]{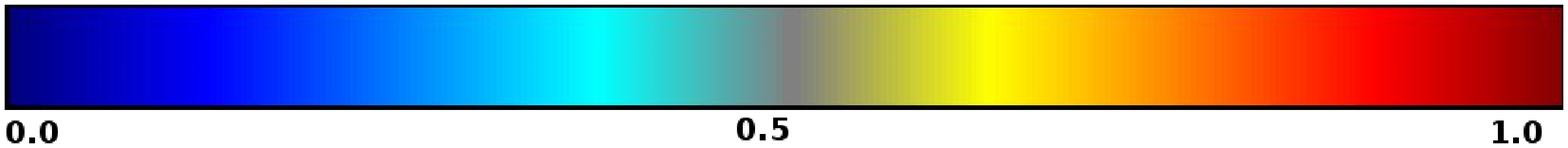}\\
\includegraphics[width=0.45\textwidth,angle=-90]{f4k.eps}
\caption{Same as in Fig.~\ref{img_trat}, but the dependency on $\tj$ is shown,
  i.e. images and spectra are for ($\tratd,\tj$)=(20,10), (20,20), and (20,30). ~\label{img_tj}}
\end{figure}

Fig.~\ref{img_tj} illustrates the dependency of the model emission on
the electron temperature in the jet. In Fig.~\ref{img_tj} we show 
images (at the same wavelengths as in Fig.~\ref{img_trat}) 
and spectra of models with ($\tratd$,$\tj$)=(20,10),
(20,20) and (20,30). Here $\mdot$ changes by a factor
of 2 for models ($\tratd$,$\tj$)=(20,10) to ($\tratd$,$\tj$)=(20,30) 
($\mdot=7.2\times10^{-8}-3.6\times10^{-8}\mdotu$).
While the appearance of the jet
at 13 and 7mm look similar, there is a dramatic change in the model appearance
at $\lambda=$1.3 mm.  For higher jet temperatures the size of the emitting
region decreases. This is because for higher jet temperatures the mass
accretion need to be lower (at a fixed 1.3 mm flux level) and for the fixed $\tratd$ the
accretion flow becomes less opaque.  Interestingly, the level of NIR emission
is sensitive to the jet temperature. Also, the spectral slope in the X-ray
band changes from $\Gamma=-0.8$ to $\Gamma=-1.5$ for models with $\tj=10$ to 30.   
The spectral slope change is due to the change of the energy of the seed
photons that undergo IC scattering in the disk. 

\subsection{Inclination angle $i$ dependence}\label{sub:3}

\begin{figure}
\includegraphics[width=0.16\textwidth]{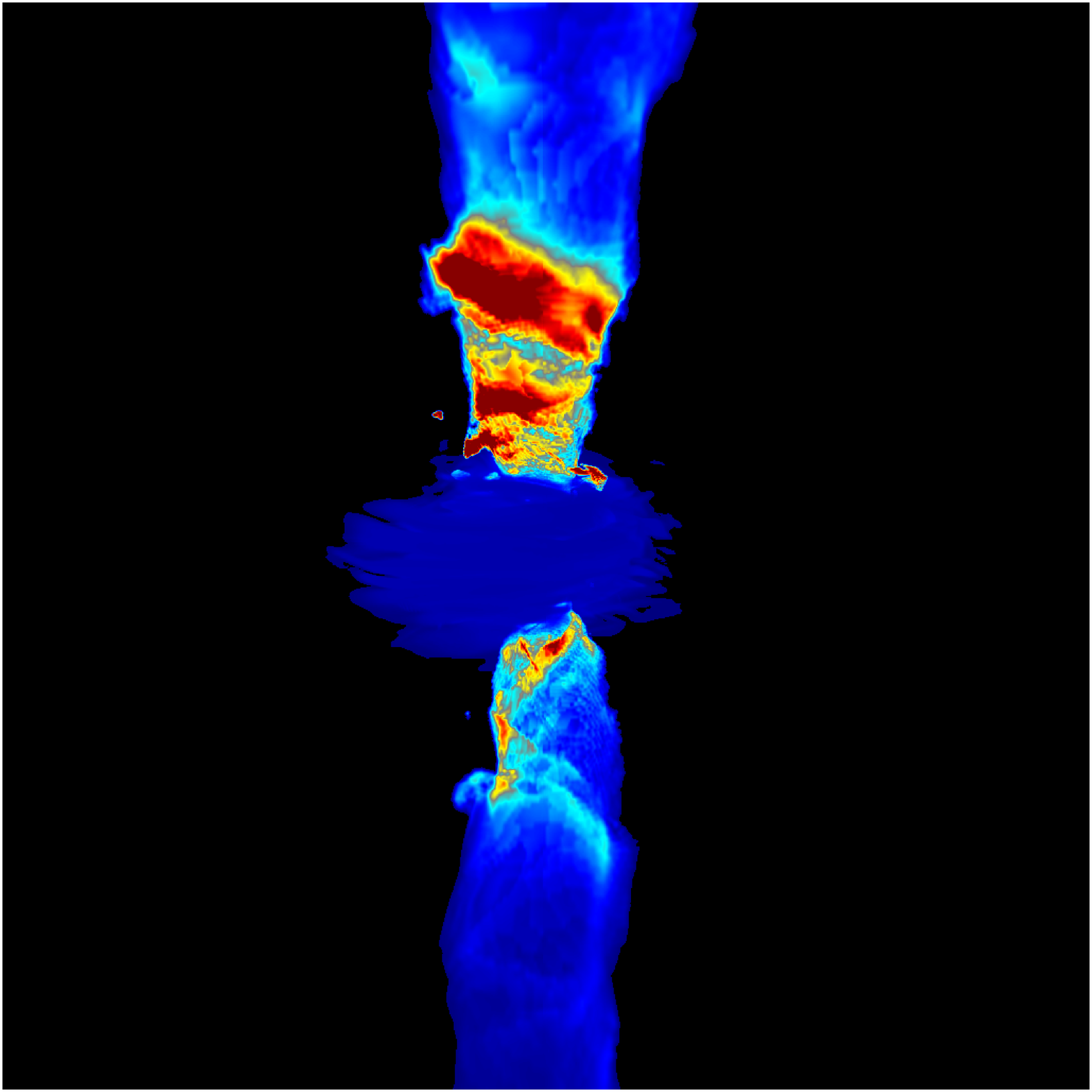}
\includegraphics[width=0.16\textwidth]{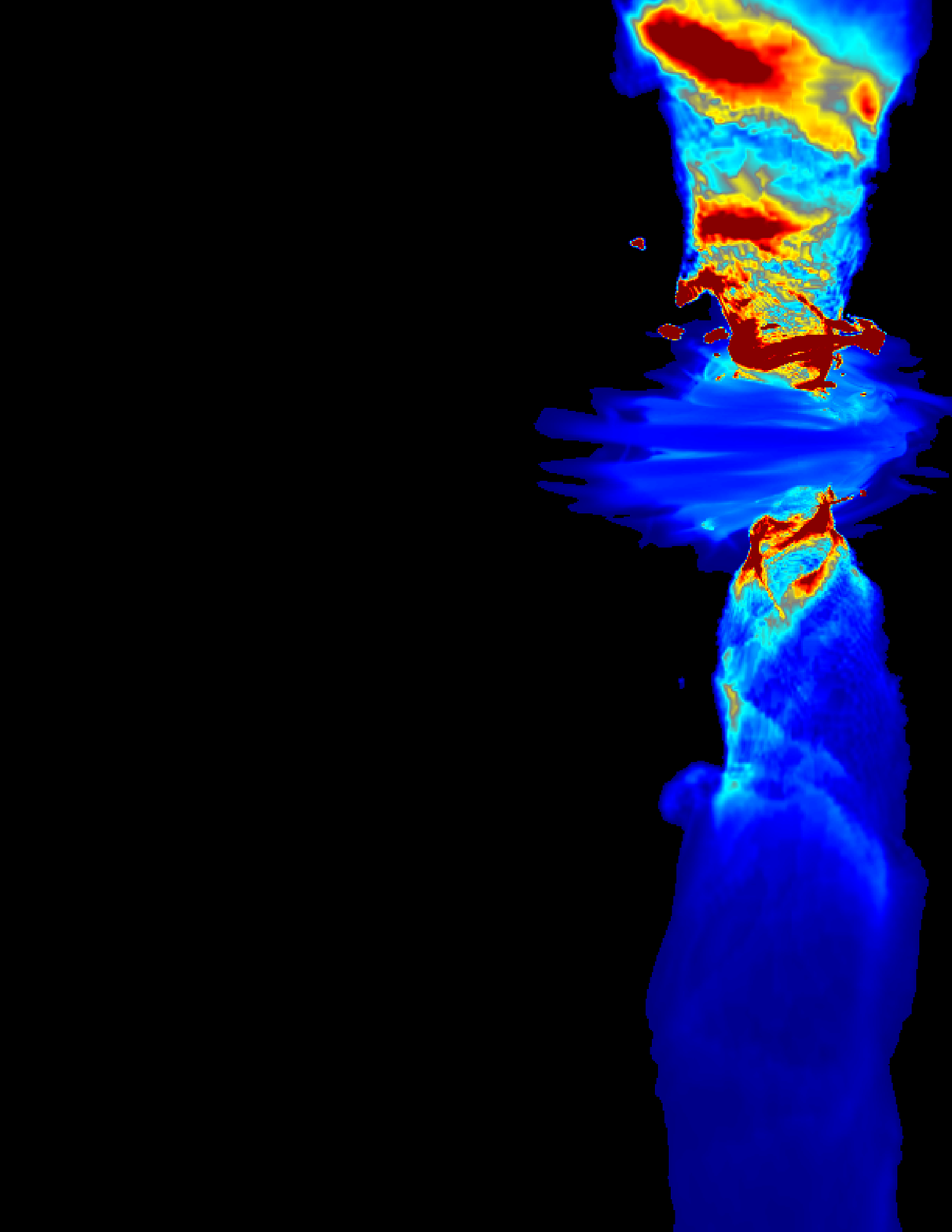}
\includegraphics[width=0.16\textwidth]{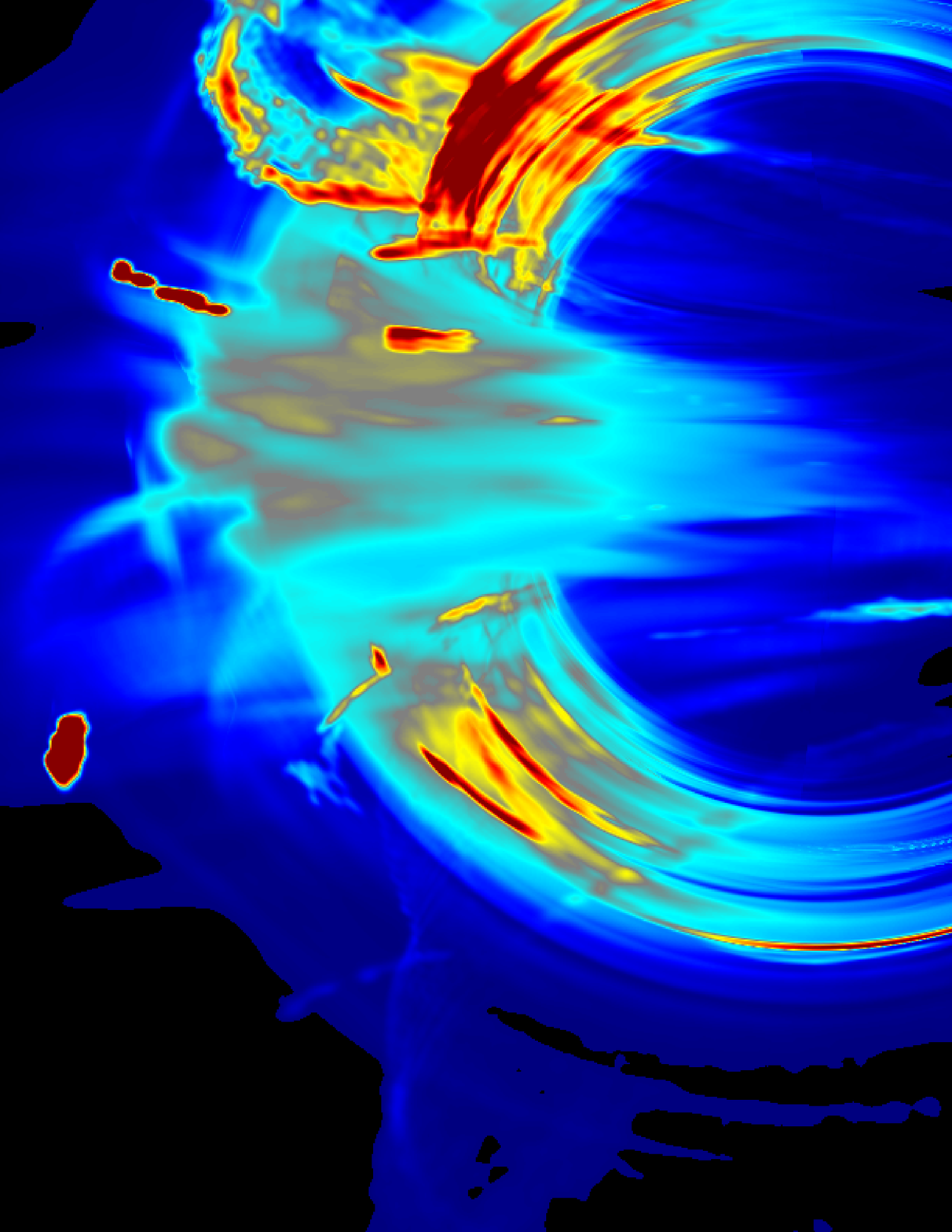}\\
\includegraphics[width=0.16\textwidth]{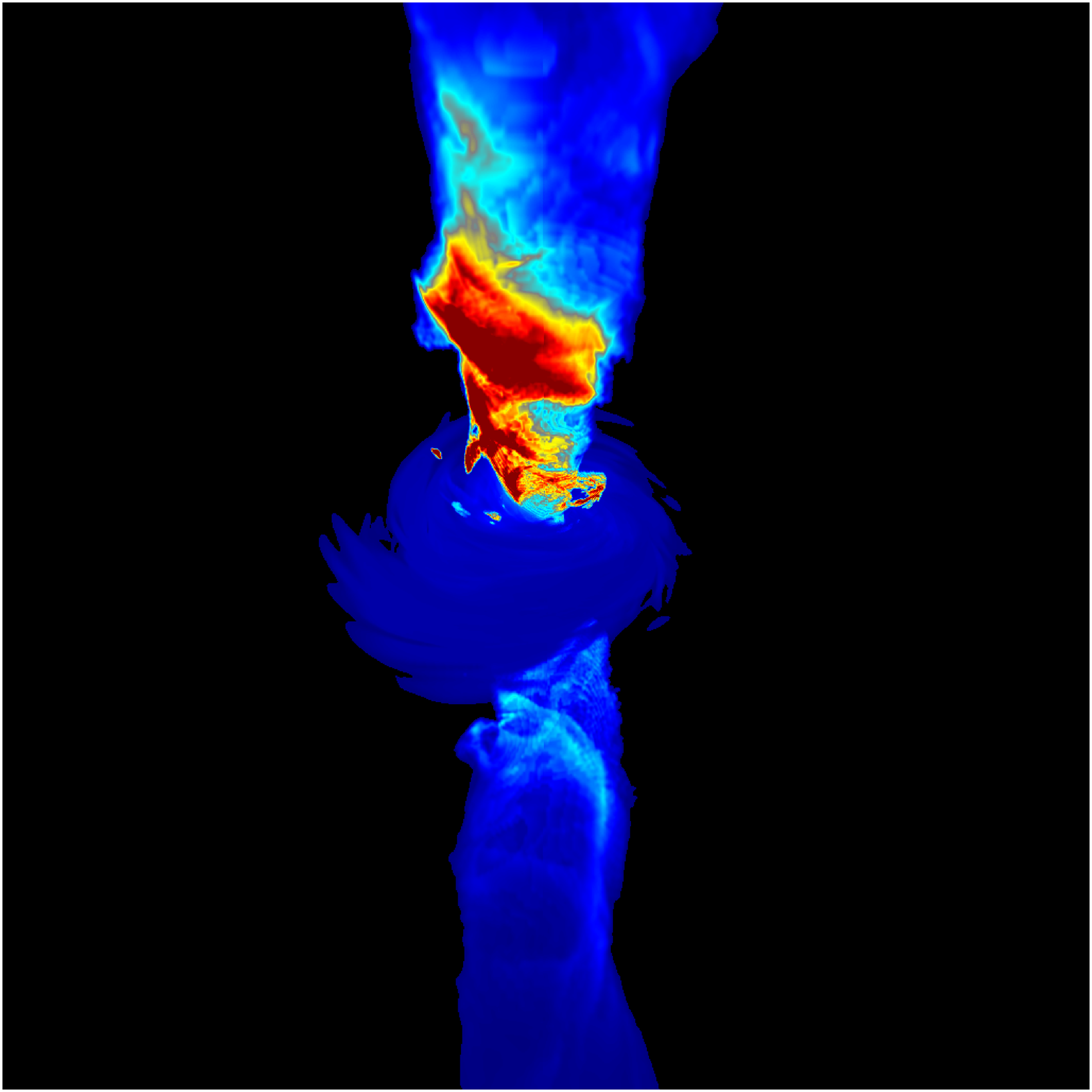}
\includegraphics[width=0.16\textwidth]{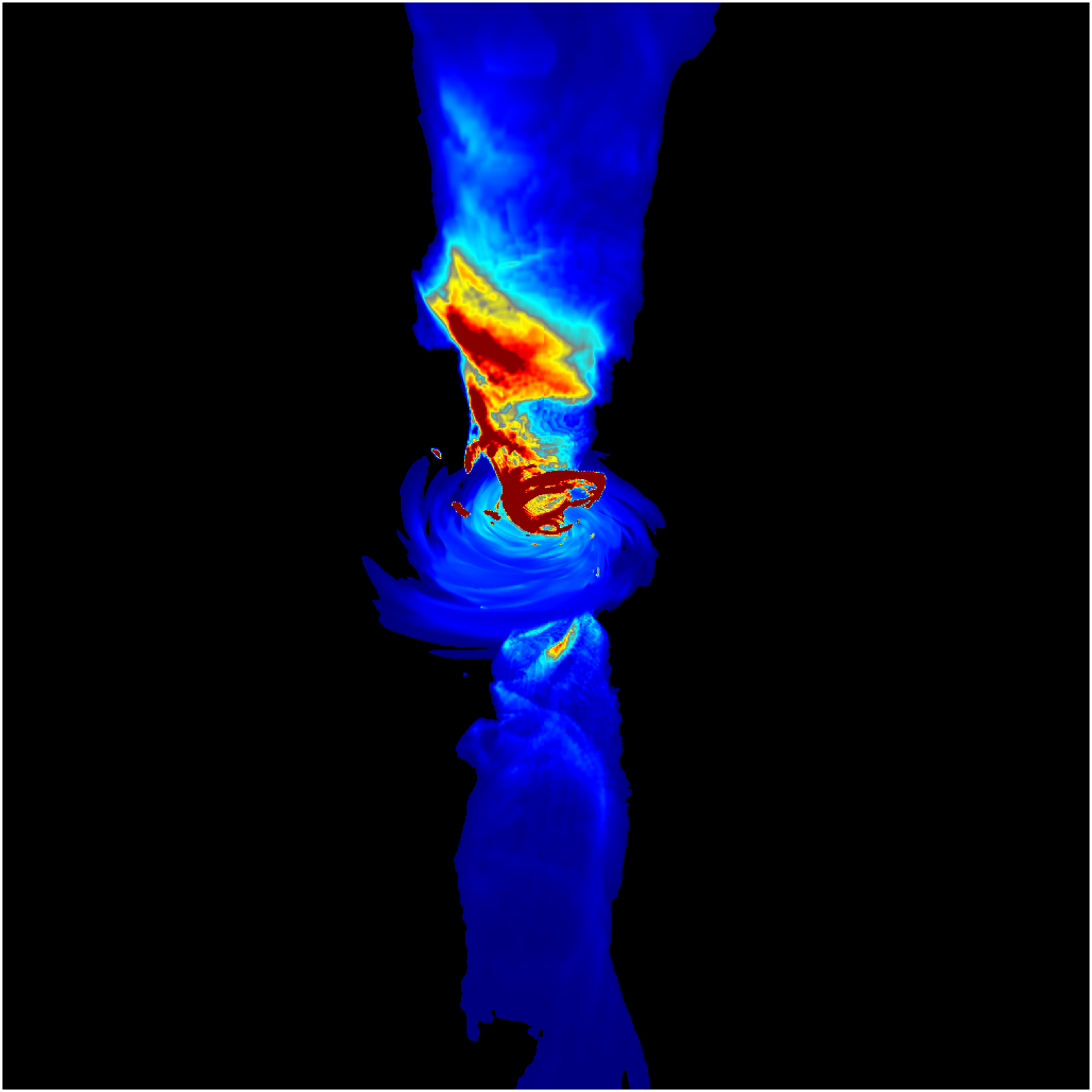}
\includegraphics[width=0.16\textwidth]{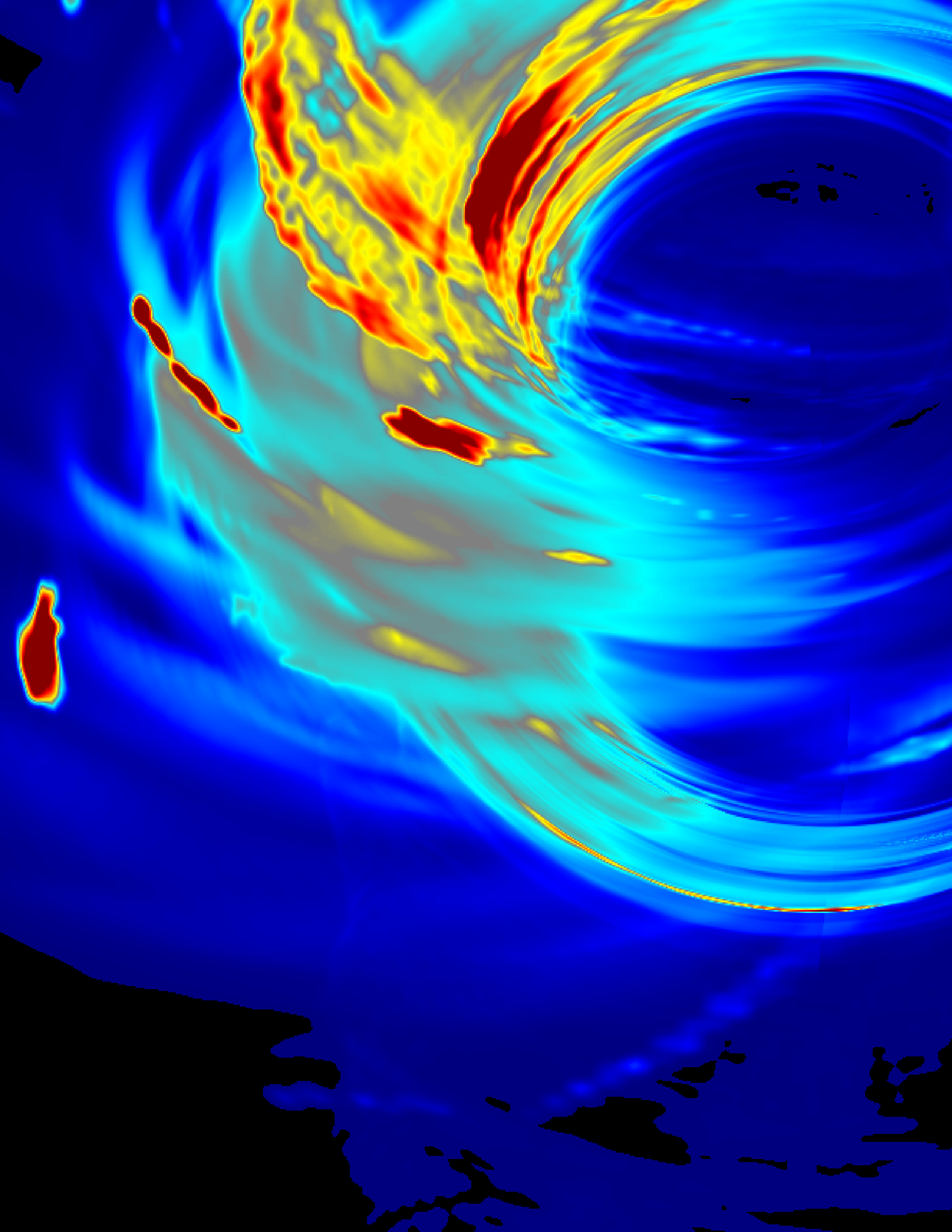}\\
\includegraphics[width=0.16\textwidth]{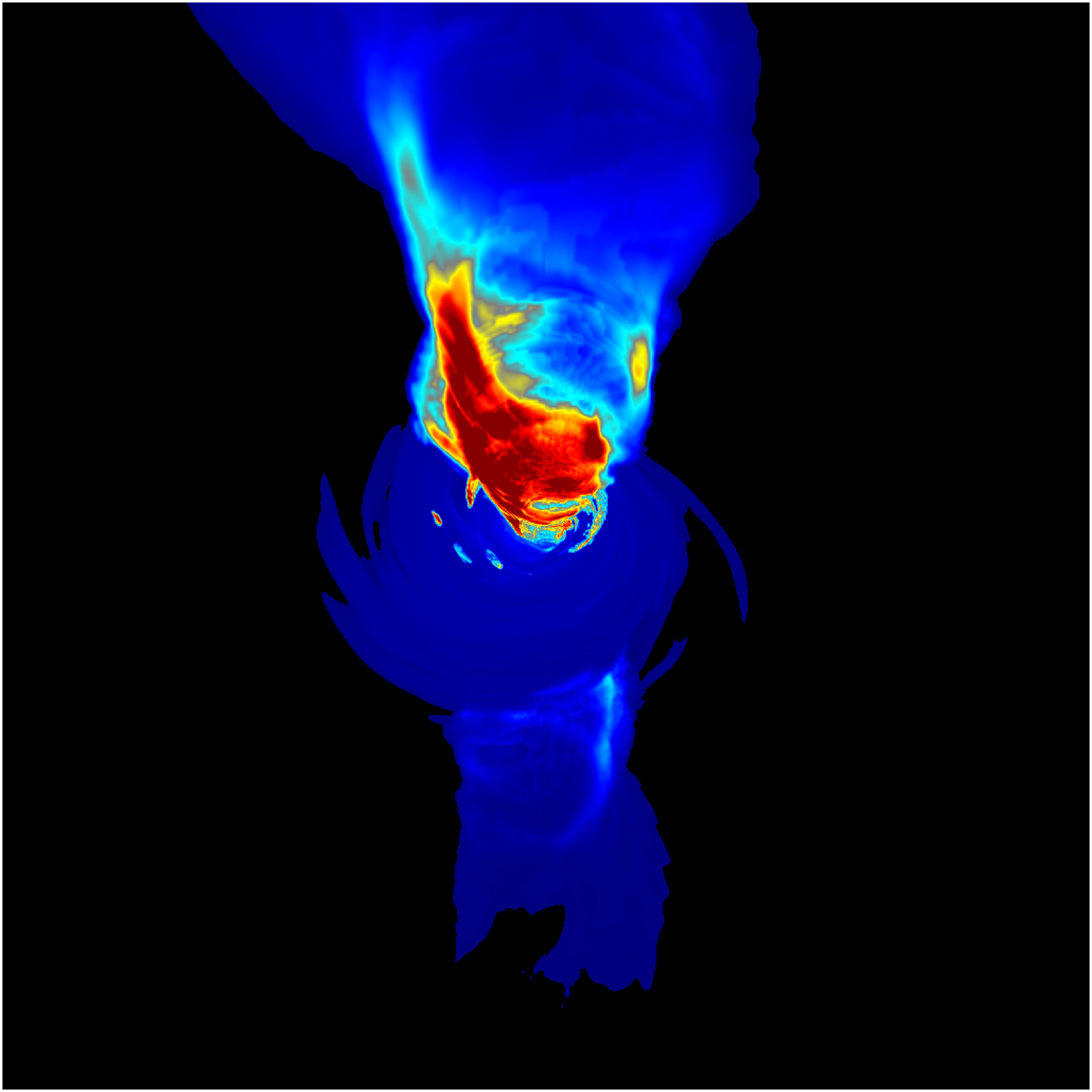}
\includegraphics[width=0.16\textwidth]{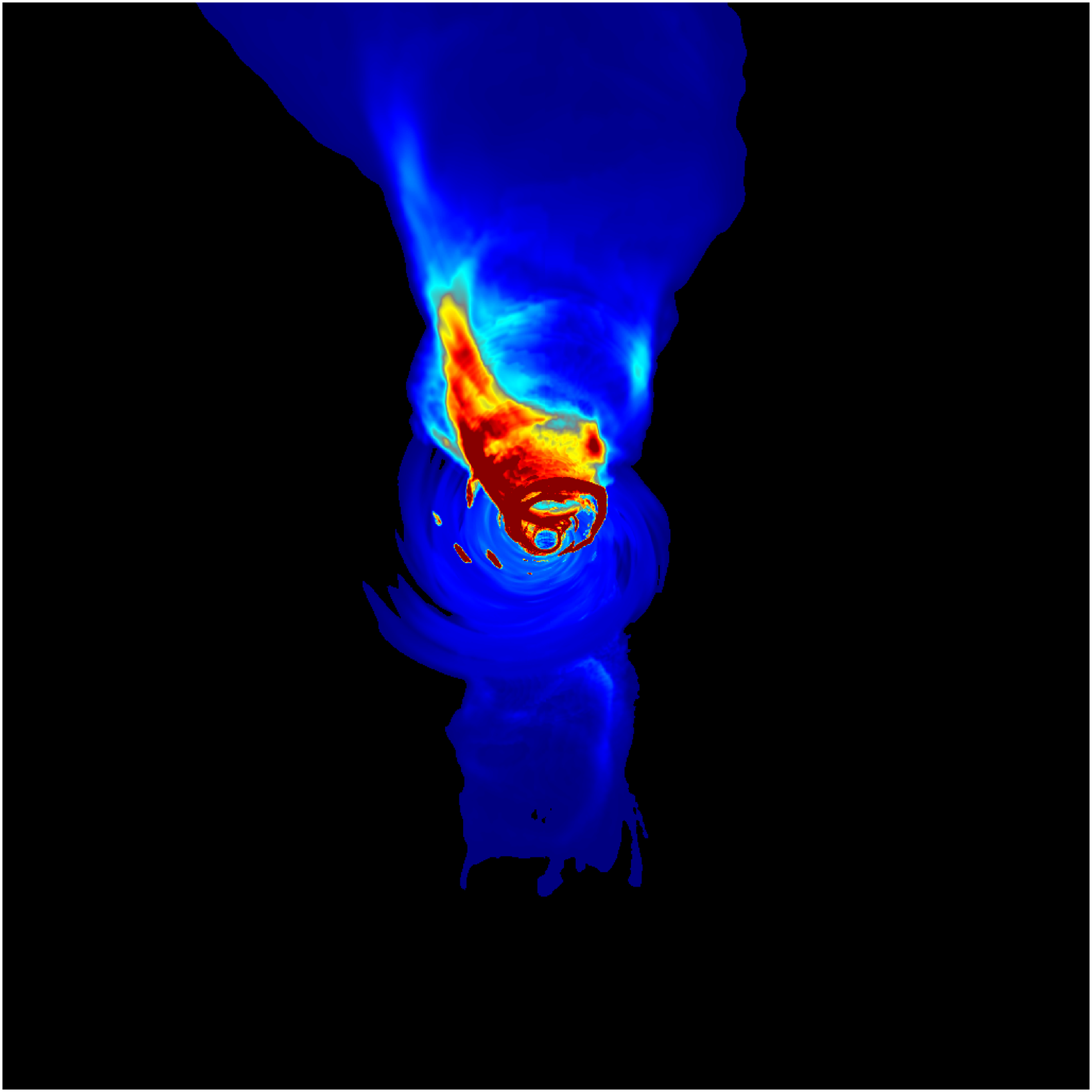}
\includegraphics[width=0.16\textwidth]{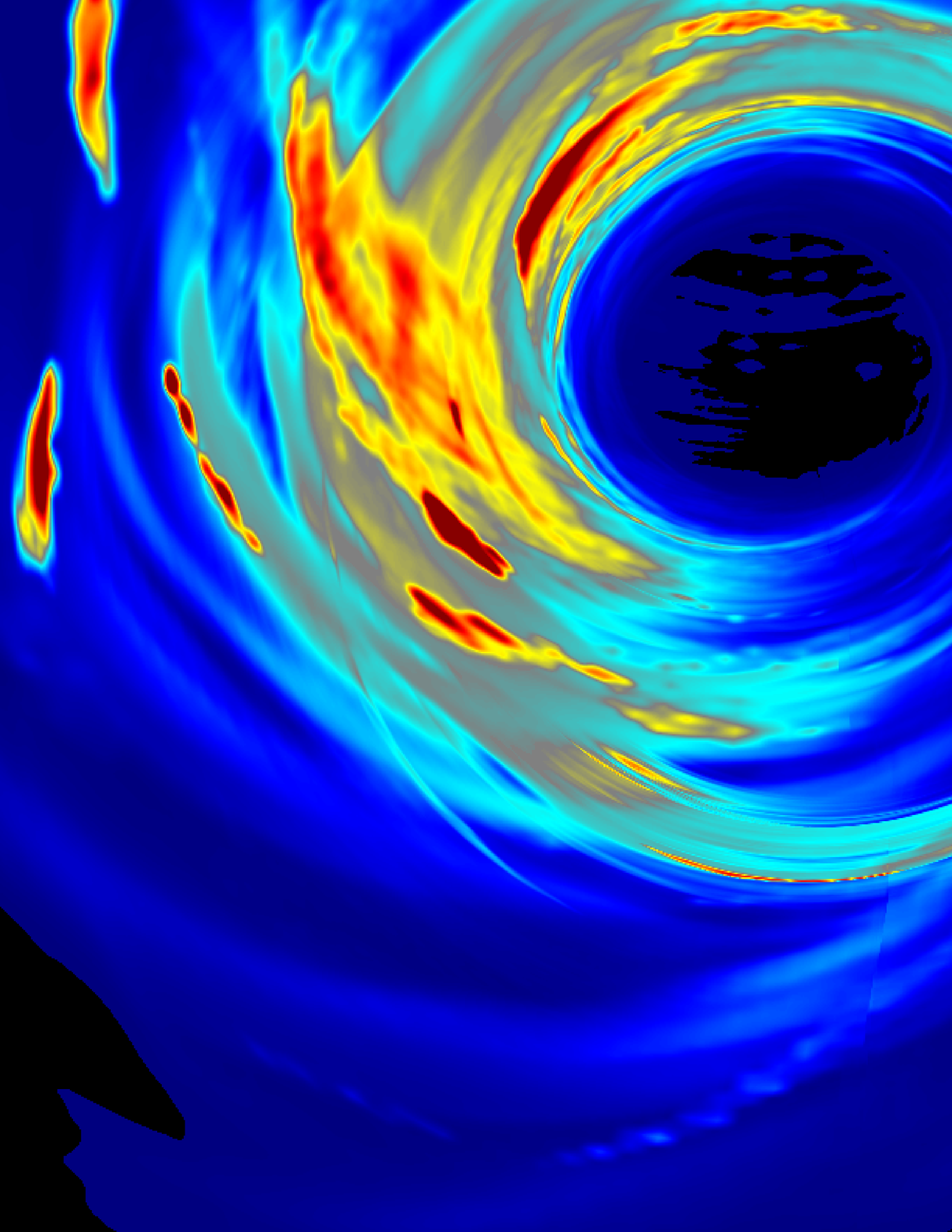}\\
\includegraphics[width=0.49\textwidth]{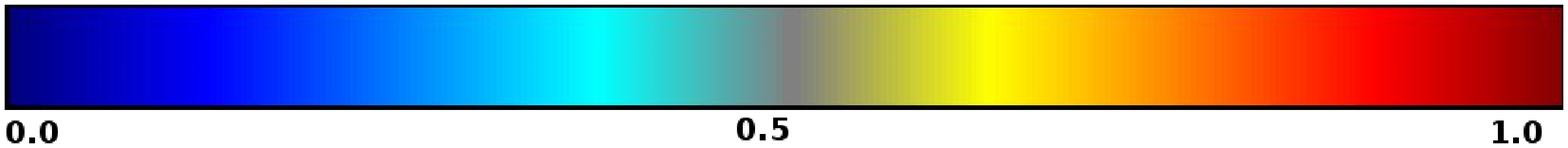}\\
\includegraphics[width=0.45\textwidth,angle=-90]{f5k.eps}
\caption{Same as in Fig.~\ref{img_trat}, but the dependency on the viewing
  angle $i$ is shown, i.e. images and spectra are for ($\tratd,\tj$)=(20,20) 
  view at  $i=90\degr$, $60\degr$, and $30\degr$.~\label{img_i}}
\end{figure}

In Fig.~\ref{img_i} (same as in Figs.~\ref{img_trat} and~\ref{img_tj})
we shows images and spectra depending on $i$. The model shown in the figure
has ($\tratd,\tj$)=(20,20), i.e., it possesses a relatively bright jet in
comparison to the disk emission. Notice, that the $\mdot$ used to normalize
the model remains roughly constant for various $i$ ($\mdot=4.6\times10^{-8}-4.1\times10^{-8}\mdotu$).

The emission at $\lambda=$13mm is predominantly produced
by the jet sheath that is much denser in comparison to the jet spine
plasma. As a consequence, the jet emission is edge-brightened when viewed at
closer to face-on inclination angles. The edge brightening is due to 
a relativistic Doppler effect. At $i=30\degr$, the
effect is observed at both, $\lambda$=7 and 13mm. At all inclination angles
the size of the emission decreases towards shorter wavelengths.

At $\lambda=$1.3mm, the emission comes from the immediate vicinity of
the black hole horizon. The images are complex but they show a common
feature: the black hole shadow is obvious at any inclination angle. It
is more pronounced compared to models with a bright disk (see e.g. Fig.~\ref{img_trat}), 
because the disk emission is more extended, due to a higher $\tratd$ and $\mdot$.

The radio SED has a spectral slope $\alpha_\nu=0.3$ and is nearly
independent of $i$. This is in contradiction to the analytical finding
of Falcke (1996), where the slope was predicted to become flatter
with smaller inclination angle. However, this effect is mainly due to
differential Doppler beaming and in the present model, the bulk flow
in the jet sheath is only very mildly relativistic.

Interestingly, the X-ray to millimeter ratio increases by about 1.5
orders of magnitude for $i$ changing from $30\degr$ to $90\degr$
and hence could be a potential inclination indicator.
The slope of the X-ray emission is constant when varying the inclination
angle.

\subsection{Modeling Sgr~A*}\label{sub:4}

\subsubsection{Spectra}

\begin{figure}[h!]
\includegraphics[width=0.5\textwidth,angle=-90.0]{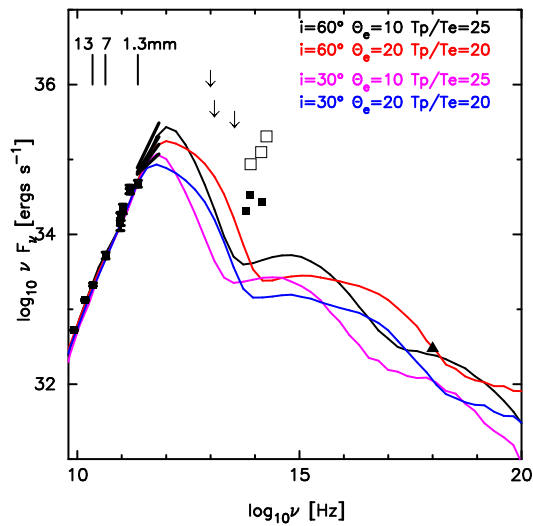}
\caption{Full SED (including synchrotron and Compton emission) of models 
  which are roughly consistent with the broadband observations of Sgr~A*
  (models \# 20, 24, 35, and 39, see Table~\ref{tab:1}.). 
  The observational data points and upper limits are
  taken from Falcke et al.(1998), An et al. (2005), Marrone et al. (2006), 
  Melia \& Falcke (2001), \citet{doeleman:2008}, Schoedel et al. (2011), 
  and the X-ray luminosity of the inner accretion flow is from 
  \citet{chandra:2013}.\label{sed:sgra}}
\end{figure}

We compare SEDs from all 45 models (Table~\ref{tab:1}) to
a selection of the Sgr~A* observational data
points collected during various observational campaigns (for references to the
observations, see caption of Fig.~\ref{sed:sgra}). We average synthetic
SEDs in time over the duration of about 3 hours to mimic the quiescent emission of 
Sgr~A*. A model should reproduce the nearly flat radio spectrum
($\alpha_\nu\approx0.3$, \citealt{falcke:1998}). Recent Chandra
observations indicate that the inner accretion flow 
should contribute at a level of $\sim$ 10\% to the total quiescent
X-ray luminosity, i.e. near the black hole the accretion flow X-ray luminosity should not
exceed $L_X \approx 4\times10^{32} {\rm ergs s^{-1}}$ \citep{chandra:2013}. 
As shown above, the X-ray emission produced by models strongly depends on the
model free parameters. The low X-ray luminosity should
in principle easily rule out some of the models.

Indeed, we have found that models with $i=90\degr$ and/or relatively
high jet temperatures ($\tj\ge30$) are too bright in the X-ray
band. All models with $\tratd=5-15$ show steep radio spectral slopes
($\alpha_\nu \sim 1$) independently of $i$ and $\tj$ and they too are
ruled out from further considerations.  There are four models where
the spectra are roughly consistent with all observed data points
between $\lambda=13$ mm and X-rays (excluding NIR energies at which the
source is constantly flaring).  These models have jets with
$\tj=10-20$ and relatively cooler disks with $\tratd=20-25$, (models
\#20, 24, 35 and 39).  Fig.~\ref{sed:sgra} displays time-averaged
spectra produced by the four models overplotted with the observations
of Sgr~A*.  Models that underpredict the observed variable component
of the X-ray luminosity produced by inner accretion flow (\#20, 35 and
39) are consistent with the data because they could be made to fit the NIR
and X-ray fluxes with the addition of a small non-thermal power-law component
in the electron distribution function.
The X-ray spectral slope of the
model SED is between $\Gamma=-1.8$ (model \#24) and $\Gamma=-1.4$
(model \#20). The mass accretion rates for the four chosen models are
in the range between $\mdot \approx 4 \times 10^{-8} - 9\times10^{-8}
\mdotu$, and they are fully consistent with the model-dependent
$\mdot$ imposed by the radio source polarization observations
(\citealt{bower:2005},\citealt{marrone:2007}).

\subsubsection{Sizes}

We have studied in greater detail images of the four models with
SEDs that are roughly consistent with observations, i.e. models \#20, 24, 35
and 39. We study sizes of images that
are time-averaged over 3 h similar to the SEDs shown in Fig.~\ref{sed:sgra}.
Sgr~A* has size estimates for
$\lambda=$1.3mm \citep{doeleman:2008} and for $\lambda=3.5{\rm mm}-6{\rm cm}$
(\citealt{bower:2006}, \citealt{shen:2005}).
At these wavelengths 
the source angular broadening by the interstellar electron scattering cannot
be neglected. To simulate the angular broadening we convolve the intrinsic
intensity maps with an elliptical Gaussian function parameterized by
\begin{subequations}\label{eq:gauss}
\begin{align}
FWHM_{maj}=1.309 (\frac{\lambda^2}{1cm}) \,\,\, {\rm mas}\\
FWHM_{min}=0.64  (\frac{\lambda^2}{1cm}) \,\,\, {\rm mas}
\end{align}
\end{subequations}
The position angle of the major axis of the scattering Gaussian is
$PA=78\degr$ measured E of N on the sky and the FWHMs are taken from
long wavelengths observations by \citet{bower:2004} and ~\citet{bower:2006}. 
We compare the model sizes to observational data at
$\lambda=$1.3, 3.5, 7 and 13mm. The radiation at $\lambda\ge13$mm cannot be
modeled in our set-up because it is emitted beyond the adopted outer
boundary of our model. Overall, we find that some 
orientations are consistent with the measured size of Sgr~A*.

\begin{figure}[h!]
\includegraphics[width=0.55\textwidth,angle=0]{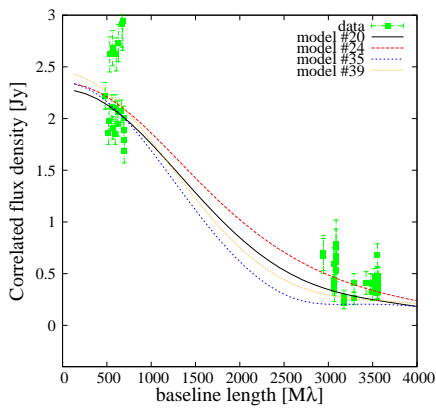}
\caption{Visibility profiles for models where the SEDs are shown in
  Fig.~\ref{sed:sgra} (i.e. models \# 20, 24, 35, and 39). Theoretical visibility
  amplitudes are computed along the E-W baselines which is roughly the the
  direction of baselines for which the 1.3mm-VLBI measurements has been done.
  The visibility data points from three days of observations are 
  from \citet{fish:2011}.}\label{fig:vis_prof}
\end{figure}

First, we check if the size of the emitting region at $\lambda$=1.3mm is
consistent with the mm-VLBI measurements.  \citet{doeleman:2008} measured and
fit the observed Sgr~A* visibilities with a symmetric Gaussian function which
correspond to the size of the source of $FWHM_{symm. Gaussian}=43^{+14}_{-8}
{\rm \mu as}$ (size estimate including angular broadening by scattering with a
FWHM=22${\rm \mu as}$ symmetric Gaussian). 
Our scattering Gaussian function and the source models are strongly asymmetric.
We therefore turn to a direct comparison of the model
to the visibility data obtained by 1.3 mm-VLBI observations. In
particular, we use data obtained by \citet{fish:2011}. 
We compute the theoretical visibility amplitudes (Fourier transformations) 
of the time-averaged, scatter-broadened model images.
Fig.~\ref{fig:vis_prof} shows model 20, 24, 35, and 39 visibility profiles
along the E-W direction which roughly corresponds to the \citet{doeleman:2008}
and \citet{fish:2011} orientation of the VLBI baselines.
Overall, the models with $i=60\degr$ tend to show better alignment with the observed
data points in comparison to models with $i=30\degr$ which seem to be too
extended to fit the data (because $FWHM_{\rm uv}\sim 1/FWHM_{\rm xy}$). 

\begin{figure*}[ht]
\includegraphics[width=0.38\textwidth,angle=-90]{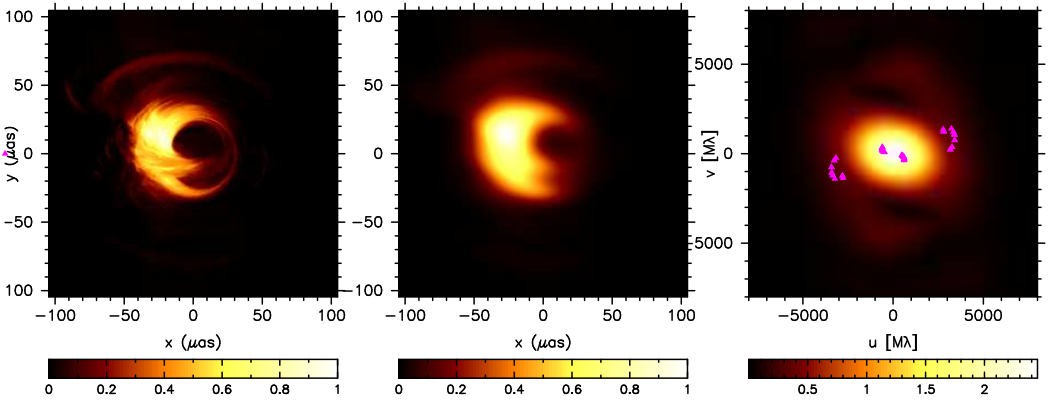}
\caption{ Intrinsic image, scatter-broadened image, and visibility amplitude
  distribution for model 24 at $\lambda=1.3mm$. Images are time-averaged
  (over $\Delta t\approx 3h$) and the color intensity
  indicates the intensity of radiation normalized to unity (linear scale).
  The visibility amplitudes are in units of Jansky.  The visibility u-v tracks are from
  \citet{fish:2011}.}\label{vis24}
\end{figure*}

The best fit model is \#24. Fig.~\ref{vis24} shows the model
image, scatter-broadened image, and visibility amplitude for the black
hole spin PA $\xi=0\degr$ with respect to the N-E
direction. The visibility amplitudes are overplotted with the
mm-VLBI u-v tracks from \citet{fish:2011}. The time-averaged image of
the model is composed of a ring, crescent-like shape, and a spot-like
emission which constitutes the image centroid. The ring-like shape
 is produced by a gravitationally lensed image of the accretion
flow and the crescent-like "tongue" is an emission feature produced by
the approaching side of the disk, which sweeps across the near side of the
black hole.  The spot-like emission is
produced by the foot-point of the isothermal jet. The structure is
smeared by the scattering effects, but the shadow of the black hole
horizon is clearly visible.  The black hole shadow is detectable in
the visibility space as two minima located at baselines nearly
orthogonal to the currently available ones.

\begin{figure}[h!]
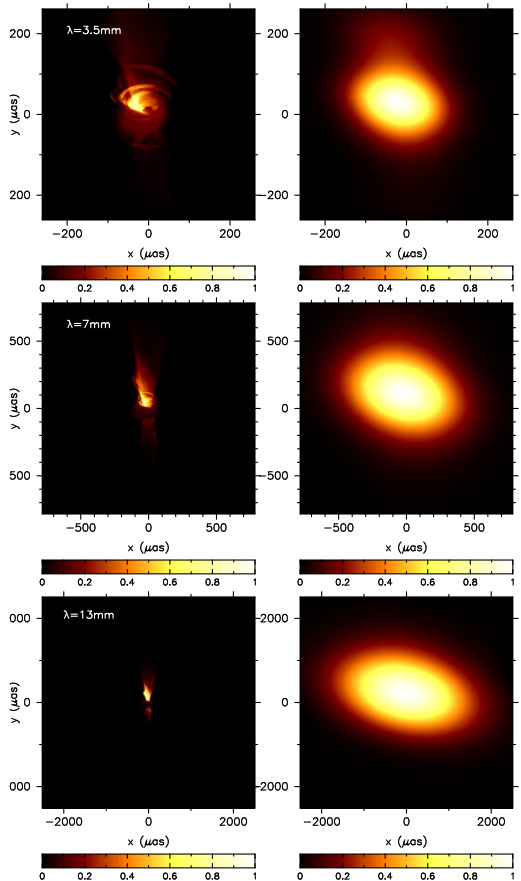

\includegraphics[width=0.28\textwidth,angle=-90]{f9a.ps}\\
\includegraphics[width=0.28\textwidth,angle=-90]{f9b.ps}\\
\includegraphics[width=0.28\textwidth,angle=-90]{f9c.ps}
\caption{Time-averaged images of model \#24 at $\lambda=$13, 7 and 3.5mm. 
  Color intensity codes the intensity of radiation normalized to unity (linear scale).
  Left panel: intrinsic intensity distribution, right panel: intensity
  distribution convolved with a asymmetric Gaussian function to simulate the
  effect of angular-broadening by scattering of radio waves 
  on the free electrons along the line of sight.\label{mmimgavg}}
\end{figure}

What does model \#24 look like at $\lambda>1.3 {\rm mm}$? 
Fig.~\ref{mmimgavg} displays
time-averaged (over $\Delta t=$3 hours, same as 1.3mm images) images of the
model \#24 at $\lambda=3.5, 7$ and 13 mm and a $\xi=0 \degr$. The left
panels are the intrinsic emission maps and the rightmost panels show the
images after convolution with the asymmetric scattering Gaussian
(Equations~\ref{eq:gauss}). In the right panel in Fig.~\ref{mmimgavg}, at
$\lambda=$3.5 mm a weak jet feature is
visible. Interestingly, in this model the shadow of the black hole is
already visible in the non-scattered images at this frequency.
At $\lambda=$7 mm and 13 mm, the source looks, as expected,
Gaussian-like despite the underlying jet-structures.

\begin{figure}
\includegraphics[width=0.45\textwidth,angle=-90]{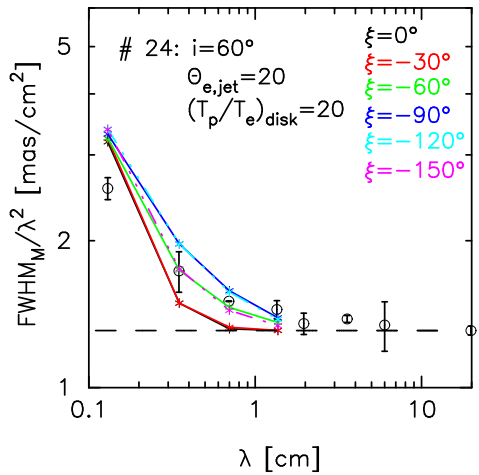}
\caption{ Wavelength-size relation for model 24 oriented at
  various position angles with respect to the NE direction. The colored lines
  correspond to various position angles of the black hole spin on the sky,
  $\xi$. Size for $\lambda>13$mm are not modeled because they fall
  outside our simulation domain. All observational data taken from Bower et al.,
  except one point at 1.3mm which is taken from Doeleman et al. (2008), which is a fit
  to the symmetric Gaussian function (due to a poor uv coverage at 1.3mm).  At
  1.3 mm the average of major and minor axis of the source are in agreement
  with the measured size. Horizontal dashed line corresponds to
  $FWHM_{maj}=1.309 \lambda^2$.}\label{jetsize}
\end{figure}

For $\lambda$ = 3.5, 7 and 13 mm, 
we measure the size of the emitting region by fitting an
asymmetric Gaussian function to the modeled scatter-broadened images (as those
in Fig.~\ref{mmimgavg}). Fig.~\ref{jetsize} shows the radio source sizes
(major FWHM) at $\lambda=0.1-13$ mm.
The size-wavelength relation might be altered when non-thermal particle tails
are added to the distribution functions (e.g. \citealt{ozel:2000}). 
At $\lambda=$1.3mm the data point is produced
by the symmetric Gaussian fit to the visibility data so the discrepancy
between the model and the data can be ignored.  At $\lambda >$ 13mm, the
model and the source follows the $\lambda^2$-law 
expected from scattering. For intermediate
wavelengths ($\lambda$ = 3.5, 7, and 13 mm) the sizes of the model major axis 
are following the same dependency as the data points but also depend on the
chosen position angle of the jet axis (or black hole spin) on the sky.
Not only is the model consistent with the data (given the uncertainties
in the electron distribution function) 
but it is interesting that the size-wavelength relation contains
information about the position angle of the jet, and therefore the black hole
spin axis, on the sky.

\begin{figure}
\includegraphics[width=0.5\textwidth]{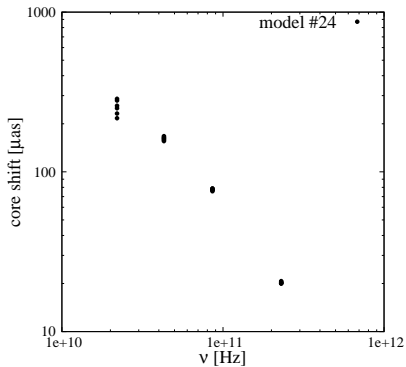}
\caption{Position of the model \#24 image centroid as a function of
  frequency (core-shift).}\label{jetcore}
\end{figure}

Finally, it is worth mentioning that models with a bright jet, like the one
considered in this subsection (model \#24) predict a significant shift of the
image centroid as a function of wavelength (so called
core-shift). Fig.~\ref{jetcore} shows the model predictions for the emission
centroid position as a function of frequency. For $\lambda>$3.5mm the core
shift is roughly proportional to the wavelength, which is consistent with
predictions from the analytical jet models \citep{falcke:1999}.  A source shift of about 300
${\rm \mu as}$ is predicted by the model when moving from 1.3 to 13mm
observations.  Measurements of the core shift would lead to the constraints of
the position angle of an eventual jet on the sky and would put an independent
constraint on the models.

\section{Summary}\label{sec:diss}

We have shown broadband spectra ($\lambda$=13 mm - $\gamma$-rays) and radio
images ($\lambda$=13, 7, 3.5, and 1.3mm) of GRMHD radiatively
inefficient accretion flow models for various electron temperature
prescriptions in inflowing and outflowing plasma of an underluminous
black hole. We have found that the assumed prescription for the electron temperature 
makes large qualitative difference in the model appearance and emission. The
presented results based on the 3-D models are broadly consistent with
those obtained from 2-D models in \citet{moscibrodzka:2013}.

It is possible to consider more sophisticated models for the electrons temperatures by,
for instance, assuming different electron-proton coupling constants in
the disk and in the jet with a natural transition controlled by, e.g.,
the magnetization of the plasma.  Even more realistic models should
include evolving the electron energy equation self-consistently with
the plasma dynamics. Future RIAF models 
should also include, for example,
the effects of pressure anisotropy and thermal conduction, which
may naturally lead to different electron distribution functions
in the turbulent accretion flow and in the jet.

We have scaled our models to Sgr~A* and we can account for
most of its observed characteristics: flat radio spectrum, 
increase of the source size with wavelength and weak X-ray emission. 
Models where the radio
emission is dominated by emission from the jet fit the data best. 
In particular our present best (bright jet) model, in contrast to the "best-bet" model from
\citet{moscibrodzka:2009}, produces X-ray luminosity that is consistent
with the new X-ray luminosities provided by very long exposure 
Chandra observations \citep{chandra:2013}.

Sgr~A* is strongly scatter-broadened at radio wavelengths and hence
the orientation of the system and detailed structure cannot be derived
directly from VLBI observations yet. Fortunately, the scattering
decreases with $\lambda^2$-law and at millimeter wavelengths the
source structure can be directly observed with a millimeter-VLBI
network. Nonetheless, current mm-VLBI and X-ray observations already rule out
a large fraction of the parameter range for models. 

Eventually,
improved mm-VLBI and flux monitoring observations will constrain the
geometry of the collisionless plasma in the immediate vicinity of the
black hole better. Thus, the ever more sophisticated RIAF models will
be constrained more and more tightly in the future, providing us with
an excellent opportunity to understand accretion and outflow in
low-power supermassive black holes as well as test general relativity reliably
with observations of the black hole shadow.

\begin{acknowledgements}
This work is supported by NWO Spinoza Price awarded to Heino Falcke and by
ERC Synergy Grant "BlackHoleCam: Imaging the Event Horizon of Black Holes" 
awarded to Heino Falcke, Michael Kramer and Luciano Rezzolla.
This work was supported in part by US NSF grants AST 07-09246 and AST
13-33612, and NASA grant NNX10AD03G. 
This work used XSEDE, which is supported by NSF grant ACI-1053575. 
MM would like to thank Ryuichi Kurosawa for his comments.

\end{acknowledgements}

\bibliographystyle{aa}
\bibliography{local}

\begin{thebibliography}{44}
\expandafter\ifx\csname natexlab\endcsname\relax\def\natexlab#1{#1}\fi

\bibitem[{{Beckwith} {et~al.}(2008){Beckwith}, {Hawley}, \&
  {Krolik}}]{beckwith:2008}
{Beckwith}, K., {Hawley}, J.~F., \& {Krolik}, J.~H. 2008, \apj, 678, 1180

\bibitem[{{Begelman}(2012)}]{begelman:2012}
{Begelman}, M.~C. 2012, \mnras, 420, 2912

\bibitem[{{Blandford} \& {K{\"o}nigl}(1979)}]{blandford:1979}
{Blandford}, R.~D. \& {K{\"o}nigl}, A. 1979, \apj, 232, 34

\bibitem[{{Bower} {et~al.}(2004){Bower}, {Falcke}, {Herrnstein}, {Zhao},
  {Goss}, \& {Backer}}]{bower:2004}
{Bower}, G.~C., {Falcke}, H., {Herrnstein}, R.~M., {et~al.} 2004, Science, 304,
  704

\bibitem[{{Bower} {et~al.}(2005){Bower}, {Falcke}, {Wright}, \&
  {Backer}}]{bower:2005}
{Bower}, G.~C., {Falcke}, H., {Wright}, M.~C., \& {Backer}, D.~C. 2005, \apjl,
  618, L29

\bibitem[{{Bower} {et~al.}(2006){Bower}, {Goss}, {Falcke}, {Backer}, \&
  {Lithwick}}]{bower:2006}
{Bower}, G.~C., {Goss}, W.~M., {Falcke}, H., {Backer}, D.~C., \& {Lithwick}, Y.
  2006, \apjl, 648, L127

\bibitem[{{Bower} {et~al.}(2014){Bower}, {Markoff}, {Brunthaler}, {Law},
  {Falcke}, {Maitra}, {Clavel}, {Goldwurm}, {Morris}, {Witzel}, {Meyer}, \&
  {Ghez}}]{bower:2014}
{Bower}, G.~C., {Markoff}, S., {Brunthaler}, A., {et~al.} 2014, ArXiv e-prints

\bibitem[{{Broderick} {et~al.}(2009){Broderick}, {Fish}, {Doeleman}, \&
  {Loeb}}]{broderick:2009a}
{Broderick}, A.~E., {Fish}, V.~L., {Doeleman}, S.~S., \& {Loeb}, A. 2009, \apj,
  697, 45

\bibitem[{{de Bruyn}(1976)}]{bruyn:1976}
{de Bruyn}, A.~G. 1976, \aap, 52, 439

\bibitem[{{Dexter} {et~al.}(2010){Dexter}, {Agol}, {Fragile}, \&
  {McKinney}}]{dexter:2010}
{Dexter}, J., {Agol}, E., {Fragile}, P.~C., \& {McKinney}, J.~C. 2010, \apj,
  717, 1092

\bibitem[{{Dexter} {et~al.}(2012){Dexter}, {McKinney}, \& {Agol}}]{dexter:2012}
{Dexter}, J., {McKinney}, J.~C., \& {Agol}, E. 2012, \mnras, 421, 1517

\bibitem[{{Doeleman} {et~al.}(2009){Doeleman}, {Agol}, {Backer}, {Baganoff},
  {Bower}, {Broderick}, {Fabian}, {Fish}, {Gammie}, {Ho}, {Honman},
  {Krichbaum}, {Loeb}, {Marrone}, {Reid}, {Rogers}, {Shapiro}, {Strittmatter},
  {Tilanus}, {Weintroub}, {Whitney}, {Wright}, \& {Ziurys}}]{doelemanwp:2009}
{Doeleman}, S., {Agol}, E., {Backer}, D., {et~al.} 2009, in Astronomy, Vol.
  2010, astro2010: The Astronomy and Astrophysics Decadal Survey, 68

\bibitem[{{Doeleman} {et~al.}(2008){Doeleman}, {Weintroub}, {Rogers},
  {Plambeck}, {Freund}, {Tilanus}, {Friberg}, {Ziurys}, {Moran}, {Corey},
  {Young}, {Smythe}, {Titus}, {Marrone}, {Cappallo}, {Bock}, {Bower},
  {Chamberlin}, {Davis}, {Krichbaum}, {Lamb}, {Maness}, {Niell}, {Roy},
  {Strittmatter}, {Werthimer}, {Whitney}, \& {Woody}}]{doeleman:2008}
{Doeleman}, S.~S., {Weintroub}, J., {Rogers}, A.~E.~E., {et~al.} 2008, \nat,
  455, 78

\bibitem[{{Dolence} {et~al.}(2009){Dolence}, {Gammie}, {Mo{\'s}cibrodzka}, \&
  {Leung}}]{dolence:2009}
{Dolence}, J.~C., {Gammie}, C.~F., {Mo{\'s}cibrodzka}, M., \& {Leung}, P.~K.
  2009, \apjs, 184, 387

\bibitem[{{Falcke}(1996)}]{falcke:1996}
{Falcke}, H. 1996, \apjl, 464, L67

\bibitem[{{Falcke} \& {Biermann}(1995)}]{falcke:1995}
{Falcke}, H. \& {Biermann}, P.~L. 1995, \aap, 293, 665

\bibitem[{{Falcke} \& {Biermann}(1999)}]{falcke:1999}
{Falcke}, H. \& {Biermann}, P.~L. 1999, \aap, 342, 49

\bibitem[{{Falcke} {et~al.}(1998){Falcke}, {Goss}, {Matsuo}, {Teuben}, {Zhao},
  \& {Zylka}}]{falcke:1998}
{Falcke}, H., {Goss}, W.~M., {Matsuo}, H., {et~al.} 1998, \apj, 499, 731

\bibitem[{{Falcke} {et~al.}(1993){Falcke}, {Mannheim}, \&
  {Biermann}}]{falcke:1993}
{Falcke}, H., {Mannheim}, K., \& {Biermann}, P.~L. 1993, \aap, 278, L1

\bibitem[{{Falcke} \& {Markoff}(2013)}]{falcke:2013}
{Falcke}, H. \& {Markoff}, S.~B. 2013, Classical and Quantum Gravity, 30,
  244003

\bibitem[{{Falcke} {et~al.}(2000){Falcke}, {Melia}, \& {Agol}}]{falcke:2000a}
{Falcke}, H., {Melia}, F., \& {Agol}, E. 2000, \apjl, 528, L13

\bibitem[{{Fish} {et~al.}(2011){Fish}, {Doeleman}, {Beaudoin}, {Blundell},
  {Bolin}, {Bower}, {Chamberlin}, {Freund}, {Friberg}, {Gurwell}, {Honma},
  {Inoue}, {Krichbaum}, {Lamb}, {Marrone}, {Moran}, {Oyama}, {Plambeck},
  {Primiani}, {Rogers}, {Smythe}, {SooHoo}, {Strittmatter}, {Tilanus}, {Titus},
  {Weintroub}, {Wright}, {Woody}, {Young}, \& {Ziurys}}]{fish:2011}
{Fish}, V.~L., {Doeleman}, S.~S., {Beaudoin}, C., {et~al.} 2011, \apjl, 727,
  L36

\bibitem[{{Fishbone} \& {Moncrief}(1976)}]{fishbone:1976}
{Fishbone}, L.~G. \& {Moncrief}, V. 1976, \apj, 207, 962

\bibitem[{{Genzel} {et~al.}(2010){Genzel}, {Eisenhauer}, \&
  {Gillessen}}]{genzel:2010}
{Genzel}, R., {Eisenhauer}, F., \& {Gillessen}, S. 2010, Reviews of Modern
  Physics, 82, 3121

\bibitem[{{Kamruddin} \& {Dexter}(2013)}]{kamruddin:2013}
{Kamruddin}, A.~B. \& {Dexter}, J. 2013, \mnras, 434, 765

\bibitem[{{Leung} {et~al.}(2011){Leung}, {Gammie}, \& {Noble}}]{leung:2011}
{Leung}, P.~K., {Gammie}, C.~F., \& {Noble}, S.~C. 2011, \apj, 737, 21

\bibitem[{{Li} {et~al.}(2013){Li}, {Morris}, \& {Baganoff}}]{li:2013}
{Li}, Z., {Morris}, M.~R., \& {Baganoff}, F.~K. 2013, \apj, 779, 154

\bibitem[{{Marrone} {et~al.}(2007){Marrone}, {Moran}, {Zhao}, \&
  {Rao}}]{marrone:2007}
{Marrone}, D.~P., {Moran}, J.~M., {Zhao}, J.-H., \& {Rao}, R. 2007, \apjl, 654,
  L57

\bibitem[{{McKinney}(2006)}]{mckinney:2006}
{McKinney}, J.~C. 2006, \mnras, 368, 1561

\bibitem[{{Mo{\'s}cibrodzka} \& {Falcke}(2013)}]{moscibrodzka:2013}
{Mo{\'s}cibrodzka}, M. \& {Falcke}, H. 2013, \aap, 559, L3

\bibitem[{{Mo{\'s}cibrodzka} {et~al.}(2009){Mo{\'s}cibrodzka}, {Gammie},
  {Dolence}, {Shiokawa}, \& {Leung}}]{moscibrodzka:2009}
{Mo{\'s}cibrodzka}, M., {Gammie}, C.~F., {Dolence}, J.~C., {Shiokawa}, H., \&
  {Leung}, P.~K. 2009, \apj, 706, 497

\bibitem[{{Narayan} {et~al.}(2012){Narayan}, {S{\"A} dowski}, {Penna}, \&
  {Kulkarni}}]{narayan:2012}
{Narayan}, R., {S{\"A} dowski}, A., {Penna}, R.~F., \& {Kulkarni}, A.~K. 2012,
  \mnras, 426, 3241

\bibitem[{{Narayan} {et~al.}(1995){Narayan}, {Yi}, \&
  {Mahadevan}}]{narayan:1995}
{Narayan}, R., {Yi}, I., \& {Mahadevan}, R. 1995, \nat, 374, 623

\bibitem[{{Neilsen} {et~al.}(2013){Neilsen}, {Nowak}, {Gammie}, {Dexter},
  {Markoff}, {Haggard}, {Nayakshin}, {Wang}, {Grosso}, {Porquet}, {Tomsick},
  {Degenaar}, {Fragile}, {Houck}, {Wijnands}, {Miller}, \&
  {Baganoff}}]{chandra:2013}
{Neilsen}, J., {Nowak}, M.~A., {Gammie}, C., {et~al.} 2013, \apj, 774, 42

\bibitem[{{Noble} {et~al.}(2007){Noble}, {Leung}, {Gammie}, \&
  {Book}}]{noble:2007}
{Noble}, S.~C., {Leung}, P.~K., {Gammie}, C.~F., \& {Book}, L.~G. 2007,
  Classical and Quantum Gravity, 24, 259

\bibitem[{{{\"O}zel} {et~al.}(2000){{\"O}zel}, {Psaltis}, \&
  {Narayan}}]{ozel:2000}
{{\"O}zel}, F., {Psaltis}, D., \& {Narayan}, R. 2000, \apj, 541, 234

\bibitem[{{Quataert}(1998)}]{quataert:1998}
{Quataert}, E. 1998, \apj, 500, 978

\bibitem[{{Quataert} \& {Gruzinov}(1999)}]{quataert:1999}
{Quataert}, E. \& {Gruzinov}, A. 1999, \apj, 520, 248

\bibitem[{{S{\c a}dowski} {et~al.}(2013){S{\c a}dowski}, {Narayan}, {Penna}, \&
  {Zhu}}]{sadowski:2013a}
{S{\c a}dowski}, A., {Narayan}, R., {Penna}, R., \& {Zhu}, Y. 2013, \mnras,
  436, 3856

\bibitem[{{Shcherbakov} {et~al.}(2012){Shcherbakov}, {Penna}, \&
  {McKinney}}]{roman:2012}
{Shcherbakov}, R.~V., {Penna}, R.~F., \& {McKinney}, J.~C. 2012, \apj, 755, 133

\bibitem[{{Shen} {et~al.}(2005){Shen}, {Lo}, {Liang}, {Ho}, \&
  {Zhao}}]{shen:2005}
{Shen}, Z.-Q., {Lo}, K.~Y., {Liang}, M.-C., {Ho}, P.~T.~P., \& {Zhao}, J.-H.
  2005, \nat, 438, 62

\bibitem[{{Tchekhovskoy} {et~al.}(2011){Tchekhovskoy}, {Narayan}, \&
  {McKinney}}]{sasha:2011}
{Tchekhovskoy}, A., {Narayan}, R., \& {McKinney}, J.~C. 2011, \mnras, 418, L79

\bibitem[{{Yuan} {et~al.}(2002){Yuan}, {Markoff}, \& {Falcke}}]{yuan:2002}
{Yuan}, F., {Markoff}, S., \& {Falcke}, H. 2002, \aap, 383, 854

\bibitem[{{Yuan} \& {Narayan}(2014)}]{yuan:2014}
{Yuan}, F. \& {Narayan}, R. 2014, ArXiv e-prints

\end{thebibliography}

\end{document}